\documentstyle[aps,preprint,psfig,axodraw]{revtex}
\tightenlines
\def\bild#1#2{    
        \vspace*{-5mm}
        \begin{center}
        \begin{math}
        \epsfxsize#2cm
        \epsffile{#1}
        \end{math}
        \end{center}
        }

\begin{document}
\draft
\title{PERIPHERAL NUCLEON--NUCLEON \\ PHASE SHIFTS  AND CHIRAL
SYMMETRY\footnote{Work supported in part by BMBF and DFG.}}
\author{N. Kaiser$^a$, R. Brockmann$^b$ and W. Weise$^a$}
\address{$^a$ Physik Department, Technische Universit\"{a}t M\"{u}nchen,\\
    D-85747 Garching, Germany \\
\smallskip $^b$ Institut f\"ur Kernphysik, Universit\"at Mainz, D-55099 Mainz,
Germany} 

\bigskip

\bigskip
\maketitle
\begin{abstract}
Within the one-loop approximation of baryon chiral perturbation theory we
calculate all one-pion and two-pion exchange contributions to the
nucleon-nucleon interaction. In fact we construct the elastic NN-scattering
amplitude up to and including third order in  small momenta. The phase shifts 
with orbital angular momentum  $L\geq2 $ and the mixing angles with $J\geq2$
are given parameterfree and thus allow for a detailed test of chiral symmetry 
in the two-nucleon system. We find that for the D-waves the $2\pi$-exchange
corrections are too  large as compared with empirical phase shifts, signaling 
the increasing importance of shorter range effects in lower partial waves. For 
higher partial waves, especially for G-waves, the model independent 
$2\pi$-exchange corrections bring the chiral prediction close to empirical NN 
phase shifts. We propose to use the chiral NN phase shifts with $L\geq 3$ as 
input in a future phase shift analysis. Furthermore, we compute the 
irreducible two-pion exchange NN-potentials in coordinate space. They turn out
to be of van-der-Waals type, with exponential screening of two-pion mass range.
\end{abstract}
\newpage


\section{INTRODUCTION}
The force between two nucleons is one of the fundamental problems in
nuclear physics. Experimentally,  the two-nucleon force is mapped out in
nucleon-nucleon scattering, which is a purely elastic process up to nucleon
laboratory kinetic energies of $T_{lab} = 280$ MeV, the $NN\pi^0$-threshold. 
Since both the target and the projectile have spin-1/2, there is a rich
spin-structure in NN-scattering. Besides total and differential cross sections
many independent spin-observables can be measured with polarized targets and/or
projectiles. At present a huge body of NN-scattering data exists and this data 
base will increase and improve in the near future, in particular with 
experiments to be  performed at the proton cooler facility COSY in J\"ulich.  

The theoretical interpretation of the elastic NN-scattering data is done in
terms of a phase shift analysis. The pertinent phase shifts are labeled by the
spectroscopic notation $^{2S+1}L_J$, with $S=0,1$ the total spin, $J=0,1,2 
\dots$ the total angular momentum and $L=J-S,\, J,\, J+S$ the orbital angular
momentum of the two-nucleon system. The total isospin of the NN-system
$I=0,1$ is readily determined by the Pauli exclusion principle, which demands 
$I+S+L$ to be odd, in order to have an antisymmetric wave function
in the combined isospin-spin-coordinate space. The spin-dependence of one-pion
exchange (longest range NN-interaction) gives rise to a tensor interaction
which causes a mixing of the (triplet) states with  orbital angular momentum
$L=J-1$ and $L=J+1$.  In order to quantify this effect a mixing angle 
$\epsilon_J$ is introduced for each $J$ in addition to the singlet and the
three triplet phase shifts. The empirical mixing angles $\epsilon_J$ are, 
however, small and do not exceed $7$ degrees up to the $NN\pi^0$-threshold. 
This weak mixing justifies the use of the spectroscopic notation $^{2S+1}L_J$
which presumes $L$ to be a good quantum number.  

The single (weakly) bound state, the deuteron in the $^3S_1$-channel with total
isospin $I=0$, shows up as a pole in the NN-scattering amplitude (on the 
physical Riemann-sheet of $T_{lab}$)  slightly below threshold at $T_{lab} =-
4.45$ MeV, corresponding to minus twice the deuteron binding energy
\cite{erwe}. In the $^1S_0$-channel with total isospin $I=1$ the
nucleon-nucleon potential is slightly too weak to form a bound state and a pole
on the second Riemann-sheet results \cite{taylor}, even much closer to 
threshold than the deuteron pole. Both the true bound 
state deuteron and the "anti-bound" state on the second Riemann-sheet slightly
below threshold are responsible for the extraordinarily large S-wave
NN-scattering lengths \cite{dumbr} ($a(^3S_1)= -5.42$ fm, $a(^1S_0)=23.75$ 
fm) and the rapid change of the S-wave phase shifts. In all other partial waves
the energy dependence is much more moderate. P- and D-wave phase shifts change
by at most $30^\circ$ from threshold up to the $NN\pi^0$-threshold at $T_{lab}
= 280$ MeV and F- and G-waves do not exceed values of $7^\circ$ in this energy
range \cite{arnd}. Thus with the exception of the two S-waves, the 
nucleon-nucleon interaction is actually rather weak in almost all channels.

An accurate description of the NN-scattering data and deuteron
properties has been found in terms of the one-boson-exchange model (OBE) and
more refined the Bonn-potential \cite{bonn} which includes multi-meson 
exchanges as well.  The basic ingredients of the OBE model are the one-pion
exchange, the exchange of the vector mesons $\rho(770)$ and
$\omega(782)$, and the exchange of a fictitious scalar isoscalar boson 
$\sigma(550)$. The latter may be viewed as an effective description of 
correlated two-pion exchange in the isoscalar scalar channel. Such 
an interpretation of the $\sigma(550)$ comes from the dispersion-theoretical 
studies of the two-pion exchange \cite{disp}, which use the empirical 
information from the crossed $\pi N$-scattering channel. The coupling
constants of the vector and scalar mesons to the nucleon are adjustable 
parameters. For each meson-nucleon vertex a "form factor" of monopole-type 
with adjustable cut-off in the range 1 GeV $<\Lambda<$ 2 GeV is
introduced. In the OBE model these form factors are needed to have convergent 
integrals in the Lippmann-Schwinger equation, which iterates the OBE potential
to infinite orders. Even though the meson-nucleon form factors are often 
interpreted (in analogy with the electromagnetic nucleon form factors) in terms
of nucleon substructure, they do not represent truely physical quantities. 
First, they cannot be  measured, and secondly they cannot even be defined in a
model independent way. As off-shell matrix elements they depend on the actual
form of the interpolating meson field, which is not unique for a composite
particle. Putting aside the non-observability of meson-nucleon form factors,
the OBE-model works very well. With about a dozen free parameters one
obtains accurate fits to NN-scattering data and deuteron properties
\cite{bonn}.    

In recent years there has been activity to understand the NN-force in a
more fundamental way from the symmetries of QCD, in particular from chiral
symmetry which governs low-energy strong interactions. In his seminal papers
Weinberg \cite{weinb} discussed the qualitative implications of chiral symmetry
for the two- and multi-nucleon forces. Ordonez, Ray and van Kolck \cite{kolck}
applied the chiral Lagrangian to the two-nucleon problem. Using time-ordered
perturbation theory and a Gaussian cut-off prescription to regularize loop
divergences they calculated the two-pion exchange NN-potential. In addition
more than 20 parameters related to short range NN-contact terms were 
introduced. The cut-off dependent potential was then transformed to coordinate
space and used to solve the non-relativistic Schr\"odinger equation. A good fit
was obtained for the deuteron properties and the phase shifts in the low
partial waves (S, P, D) up to kinetic energies of  $T_{lab}=100 $ MeV (partly
also up to $T_{lab} =300$ MeV) \cite{kolck}. Of course in the presence of
nearly 30 fit parameters the role of chiral symmetry in the NN-interaction 
is obscured. Furthermore, the use of a Gaussian cut-off is not consistent with
the chiral power counting scheme underlying the whole approach \cite{weinb}.  

The purpose of this work is to explore the kinematical window in energy and
angular momentum in which the NN-interaction is governed by chiral symmetry
alone. Our physical motivation comes from the simple fact that the range of the
force produced by the exchange of a hadronic state is inversely proportional to
its mass. The longest range part comes from one-pion exchange and chiral
symmetry manifests itself via the smallness of the pion mass, a feature which
derives from  the fact that the pion is an approximate Goldstone boson of
spontaneous chiral symmetry breaking in QCD. The exchange of the next heavier
hadronic state, consisting of two pions, is intimately related to $\pi
N$-scattering by crossing. In this reaction chiral symmetry leads to important 
dynamical constraints in the form of low-energy theorems \cite{alfaro}, see
also the recent work of \cite{pin}, where many low-energy $\pi N$-observables
have been  investigated in chiral perturbation theory. It is now interesting to
explore the structure of the $2\pi$-exchange NN-potential when it is produced 
by a $\pi N$-interaction with all chiral constraints built in. Naturally, one
expects such an approach to work for the peripheral NN-partial waves where
only the long and intermediate range components of the NN-force operate.

In the present work we calculate the chiral two-pion exchange based 
on the same chiral $\pi N$-Lagrangian as used in \cite{kolck}. Our methods and
aims are however different and complementary. First of all we will use
covariant perturbation theory throughout and thus have to solve the problem of
the so-called pinch singularity \cite{weinb} occuring in the planar box
diagram. Secondly, we use throughout dimensional regularization, a 
prescription to handle loop divergences which is in harmony with chiral
symmetry and the chiral power counting scheme. Thirdly, we stay within the 
systematic expansion of chiral perturbation theory and calculate in 
one-loop approximation all contributions to the NN T-matrix up to and including
third order in small external momenta. Consequently, we do not iterate to
infinite orders. As a result we find that the phase shifts with $L\geq2$ and
mixing angles with $J\geq2$ are given entirely parameterfree and thus allow 
for a rigorous test of  chiral symmetry in the NN-interaction. 
 
The paper is organized as follows. In section 2, we briefly discuss the chiral
$\pi N$-Lagrangian at next-to-leading order underlying our calculation.  
In section 3, we introduce the necessary formalism, the nucleon-nucleon
T-matrix, the projection onto partial waves and the expressions for the 
perturbative phase-shifts and mixing angles. Section 4 presents  closed form
analytical results for the one-pion and two-pion exchange contributions to the
NN T-matrix calculated in one-loop approximation. In section 5 the
parameterfree results for the peripheral phase shifts $(L\geq2)$ and mixing
angles $(J\geq2)$ are discussed. In section 6 we give explicit 
expressions for the irreducible two-pion exchange potentials in coordinate 
space and compare them with phenomenological ones. In section 7 we compare the
isoscalar central amplitude at zero momentum transfer with OBE and finally
section 8 ends with a summary. In the appendix the construction of the 
anti-symmetric NN T-matrix satisfying the Pauli exclusion principle is given.

\section{EFFECTIVE CHIRAL LAGRANGIAN}
The tool to investigate the dynamical consequences of spontaneous and explicit 
chiral symmetry breaking in QCD is the effective chiral Lagrangian. It provides
a non-linear realization of the chiral symmetry group  $SU(2)_L \times 
SU(2)_R$ using the relevant low-energy effective degrees of freedom, the
Goldstone pions  and the nucleons. The effective chiral $\pi N$-Lagrangian
takes the following general form,
\begin{equation}
{\cal L} = {\cal L}_{\pi N}^{(1)} + {\cal L}_{\pi N}^{(2)} +\dots 
\end{equation} 
where the superscript indicates the number of derivatives or small external
momenta. In the heavy baryon formulation \cite{jenk} (which essentially 
corresponds to a non-relativistic treatment of the nucleons) the leading and
next-to-leading order terms read \cite{pin,review},   
\begin{eqnarray}
{\cal L}_{\pi N}^{(1)} &=& \bar N \Big( i D_0 - {g_A\over 2} \vec \sigma \cdot
\vec u \Big) N \,, \\ 
{\cal L}_{\pi N}^{(2)} &=& \bar N \Bigg( {1 \over2 M}\vec D\cdot \vec D  +i
{g_A\over 4 M} \big \{ \vec \sigma \cdot \vec D, u_0 \big \} + 2c_1\, m_\pi^2 
( U+U^\dagger) \nonumber \\ & & +\Big(c_2-{g_A^2\over 8M}\Big)  u_0^2
+c_3\, u_\mu u^\mu  +{i\over2}\Big(c_4+{1\over 4M} \Big) \vec\sigma \cdot (\vec
u \times \vec u\,)  \Bigg) N \, .  \end{eqnarray}
Here, $D^\mu = \partial^\mu + {1\over2}[ \xi^\dagger ,\partial^\mu \xi ]$ is 
the chiral covariant derivative acting on the iso-doublet nucleon field $N$ 
and $u^\mu = i \{ \xi^\dagger ,\partial^\mu \xi \}$ is an axial vector (matrix)
quantity. The $SU(2)$-matrix  $U=\xi^2$ collects the Goldstone pion fields in
the form $U= 1 + i\vec \tau\cdot \vec \pi/f_\pi- \vec\pi^2/2f_\pi^2 + \dots$
with  the pion decay constant $f_\pi = 92.4$ MeV. The only other parameter
occuring in ${\cal L}^{(1)}_{\pi N}$ is $g_A$, the axial vector coupling
constant of the nucleon. Since we are considering here exclusively the
pion-nucleon vertex we employ the  Goldberger-Treiman relation $g_A = g_{\pi N}
f_\pi/M$ together with the empirical value of $g_{\pi N} = 13.4$
\cite{hoehler} and therefore use $g_A=1.32$ throughout. $M=939$ MeV is the
(average) nucleon mass and $m_\pi= 138$ MeV the (average) pion mass. 

The second order Lagrangian ${\cal L}_{\pi N}^{(2)}$ consists of two different
types of terms.  The first two terms in eq.(3) have coefficients $1/2M$ and
$g_A/4M$ which are fixed by Lorentz invariance \cite{jenk}. The remaining set
of chiral invariant terms are accompanied by new low-energy constants
$c_1,c_2,c_3,c_4$. These four low-energy constants have been recently
determined at one-loop order in ref.\cite{pin} from a fit to nine
(sub)threshold $\pi N$-observables. Their values are $c_1 = -0.9 \pm 0.1$, $c_2
= 3.3 \pm 0.2$, $c_3 = -5.3 \pm 0.2$, $c_4 =3.6 \pm 0.1$, all given in
GeV$^{-1}$. As shown in ref.\cite{pin} the $\Delta(1232)$-resonance makes the
dominant contribution to $c_2, c_3$ and $c_4$. The exchange of a scalar boson 
with  mass and coupling constant similar to the fictitious $\sigma(550)$ 
allows to explain $c_1$ and also part of $c_3$. Furthermore, $c_4$ receives a 
significant contribution from $\rho(770)$ vector meson exchange. Note that the
$c_i$ are much larger than the typical $1/2M$-scale in eq.(3). Thus one expects
the important $2\pi$-exchange contributions to the NN T-matrix to be the ones
proportional to $c_i$. This completes the description of the effective chiral
$\pi N$-Lagrangian.      

\section{FORMALISM}
We are considering elastic nucleon-nucleon scattering $N(\vec p\,) + N(-\vec
p\,) \to  N(\vec p\,')+N(-\vec p\,')$, where $\pm \vec p$ and $\pm \vec p\,'$
are the initial and final nucleon momenta in the center-of-mass (CM) frame with
$p =|\,\vec p\,|=|\,\vec p\,'\,|$. The momentum transfer $\vec q = \vec p\,'
-\vec p$ has the magnitude $q = |\,\vec q\,| = 2 p \sin{\theta\over2} =
p\sqrt{2(1 -z)}$. Here $\theta$ is the CM scattering angle
with $z = \cos \theta$.  In terms of the nucleon laboratory kinetic
energy $T_{lab}$, the CM momentum $p$ is given as $p=\sqrt{T_{lab} M/2}$. 
 
In the center-of-mass frame the elastic on-shell nucleon-nucleon T-matrix takes
the following general form,
\begin{eqnarray} & &{\cal T}_{NN} = V_C+ \vec \tau_1 \cdot \vec \tau_2  W_C +
\big[V_{S}+ \vec \tau_1 \cdot \vec \tau_2  W_{S} \big]\,\vec\sigma_1
\cdot \vec \sigma_2+ \big[ V_T + \vec \tau_1 \cdot \vec \tau_2 W_T 
\big]\,  \vec  \sigma_1 \cdot \vec q \, \vec \sigma_2 \cdot \vec q + \nonumber 
\\ & & \quad \quad\big[ V_{SO}+\vec\tau_1 \cdot \vec \tau_2 W_{SO} \big] \,i( 
\vec \sigma_1 +\vec  \sigma_2)\cdot (\vec q\times \vec p\,)  + \big[ V_Q+
\vec \tau_1 \cdot \vec \tau_2 W_Q \big]\,\vec\sigma_1\cdot(\vec q\times 
\vec p\,) \,\vec \sigma_2 \cdot(\vec q\times \vec p\,)\end{eqnarray}
with the ten complex functions $V_C(p,z), \dots, W_Q(p,z)$ depending on $p$ and
$z$. Here $\vec \sigma_1$ and $\vec \sigma_2$ are the spin-vectors of the two
nucleons. The subscripts refer to the Central, Spin-spin, Tensor, Spin-Orbit
and Quadratic spin-orbit components of the NN T-matrix. Each of these five
components occurs in an isoscalar $(V)$ and an isovector $(W)$ version. Note
that, for the purpose of having projection formulas as simple as possible (see
eqs.(6-9)), we do not follow the usual conventions for the decomposition of
${\cal T}_{NN}$ in all cases. In particular, we associate the tensor term with
the helicity structure $\vec \sigma_1 \cdot \vec q \,\vec \sigma_2 \cdot \vec
q$ in our decomposition. The overall sign of ${\cal T}_{NN}$ is chosen here as
it follows directly from the Feynman rules for the S-matrix, with a positive
T-matrix, positive phase-shifts and a positive scattering length corresponding
to attraction.  

We remark that eq.(4) is not just a 
non-relativistic approximation, but results from the evaluation of the fully
relativistic NN T-matrix in the center-of-mass system when reduced to 
two-component spinors. In principle the NN T-matrix consists of the
total sum of all direct diagrams and those with crossed nucleon lines. In 
practice, there is however no need to evaluate the diagrams with crossed
nucleon lines, since the Pauli exclusion principle gives a simple rule, namely
$L+S+I$ being odd, to select the allowed NN-partial waves. Therefore we
identify here ${\cal T}_{NN}$ in eq.(4) only with the sum of all direct
diagrams. In the appendix it is shown how to construct from the sum of direct
diagrams the complete anti-symmetrized form  by application of
Fierz-transformations and the substitution $z \to -z$.  

In order to compute phase shifts and mixing angles the matrix elements of
${\cal T}_{NN}$ in the  $LSJ$-basis  are needed. To obtain these one
first evaluates matrix elements in the helicity basis and then rotates with
Wigner d-functions to the $LSJ$-basis. We have followed the detailed
description in ref.\cite{erkel} and find with the relevant linear combination 
in a state with total isospin $I=0,1$,
\begin{equation} U_K= V_K+(4I-3) W_K \,\,, \qquad (K=C,S,T,SO,Q) 
\end{equation}   
the following projection formulas:

\noindent
a) Singlet matrix element with $S=0$ and $L=J$: 
\begin{equation}\langle J0J|{\cal T}_{NN}|J0J\,\rangle ={1\over2}\int_{-1}^1 dz
\Big[ U_C-3U_{S}-q^2 U_T+p^4(z^2-1) U_Q \Big] P_J(z)\,.\end{equation}
b) Triplet matrix element with $S=1$ and $L=J$:
\begin{eqnarray} & &\langle J1J|{\cal T}_{NN}|J1J\,\rangle={1\over2}\int_{-1}^1
dz \Big\{2p^2\Big[ U_{SO}-U_T+p^2z U_Q \Big]\Big(P_{J+1}(z)+P_{J-1}(z)
\Big) \nonumber \\ & & \qquad \qquad+\Big[ U_C+ U_{S}+2p^2(1+z)  U_T-
4p^2z U_{SO}-p^4(3z^2+1) U_Q \Big] P_J(z) \Big\}\,. \end{eqnarray}
c) Triplet matrix elements with $S=1$ and $L=J\pm1$:
\begin{eqnarray} & & \langle J\pm1,1J|{\cal T}_{NN}|J\pm1,1J\,\rangle={1\over2}
\int_{-1}^1 dz \Big\{2p^2\Big[ U_{SO}\pm{1\over2J+1}\Big( U_T-p^2z
U_Q\Big)\Big] P_{J}(z) \nonumber \\ & & \quad \quad+\Big[ U_C+ U_{S}+p^2
\Big( p^2(1-z^2) U_Q-2z U_{SO} \pm{2\over 2J+1} \Big( p^2 U_Q-
U_T \Big) \Big)  \Big] P_{J\pm1}(z) \Big\} \,.\end{eqnarray}
d) Triplet mixing matrix element with $S=1$, $L'=J-1$ and $L=J+1$:
\begin{eqnarray}  \langle J-1,1J|{\cal T}_{NN}|J+1,1J\,\rangle&=&{\sqrt{J+1}p^2
\over \sqrt{J}(2J+1)} \int_{-1}^1 dz \Big\{\Big(U_T-p^2 U_Q\Big) 
P_{J+1}(z)  \nonumber \\ & & +\Big[\big(2J-z(2J+1)\big) U_T+p^2z U_Q 
\Big] P_J(z) \Big\} \,. \end{eqnarray}
Here, $P_J(z)$ are ordinary Legendre polynomials of degree $J$. In accordance 
with ref.\cite{erkel} we did not use the more conventional tensor operator 
$3 \vec \sigma_1 \cdot  \hat q \,\vec \sigma_2 \cdot \hat q -\vec
\sigma_1 \cdot \vec \sigma_2$ in the decomposition of ${\cal T}_{NN}$, eq.(4),
since it leads  to unnecessarily complicated projection formulas. 

The phase shifts and mixing angles (in the standard convention of Stapp et
al. \cite{stapp}) are then given perturbatively \cite{gass} as 
\begin{equation} \delta_{LSJ} = {M^2 p \over 4\pi E}\, {\rm Re} \,\langle LSJ 
|{\cal T}_{NN} | L S J\,\rangle \,, \quad\quad\epsilon_J= {M^2p\over 4\pi
E } \, {\rm Re}\langle J-1,1J|{\cal T}_{NN}|J+1,1J\,\rangle \,,  \end{equation}
with the CM nucleon energy $E=\sqrt{M^2+p^2}$.  The kinematical prefactor comes
from the proper relativistic flux and two-body phase space factors.
Since we use chiral perturbation theory here, there is no exact but only
a perturbative unitarity. Therefore the calculation of the phase shift
according to eq.(10) is valid only as long as the difference between $ \delta$ 
and $ \sin\delta \cos\delta$ is small. This holds reasonably well for $|
\delta| < 30^\circ$; thus with respect to this constraint all NN phase shifts 
and mixing angles, with the exception of the two S-wave phase shifts, can be
calculated reliably in (chiral) perturbation theory.     

\section{ONE LOOP CALCULATION OF THE NN T-MATRIX}
In this section we present the analytical results for the NN T-matrix in 
one-loop approximation. We evaluate the contributions at second and third order
of the expansion in powers of small external momenta (here $p$, $q$ and
$m_\pi$). These contributions are divided into three classes: vertex and
propagator corrections to one-pion exchange, irreducible two-pion exchange and
iterated one-pion exchange. The latter gives rise to a non-vanishing imaginary
part in the amplitudes and thus restores unitarity (perturbatively). The
possible local NN-contact terms \cite{kolck} are not considered here, since to
the order we are working the corresponding polynomial amplitudes do not
contribute to the phase-shifts with $L\geq2$ and mixing angles with $J\geq2$. 

\subsection{ONE-PION EXCHANGE}

Fig.1 shows the one-pion exchange diagram together with (some) one-loop vertex
and propagator corrections to it.  We have evaluated all possible diagrams of
this sort and found that such one-loop corrections only contribute to mass and
coupling constant renormalization.  No pion-nucleon "form factor" is generated
by any of these diagrams, neither at second nor at third  order in the small
momentum expansion. Consequently, the one-pion exchange diagrams of Fig.1 lead
just to the following well-known  contribution to the NN T-matrix,   
\begin{equation} W_T={g_{\pi N}^2\over 4 M^2 (m_\pi^2+q^2) }\,,\end{equation}
with $g_{\pi N},\, M,\, m_\pi$ the physical values of the pion-nucleon coupling
constant, nucleon mass and pion mass. We also note that in the CM frame there
are no $1/M$-corrections at any order to eq.(11). It is an exact result of
the fully relativistic pseudovector (or pseudoscalar) $\pi NN$-vertex evaluated
in the CM frame. Finally, we remark that the pionic counter terms of ${\cal
L}^{(4)}_{\pi \pi}$ in ref.\cite{leut} and the on-shell (single) nucleon
counter terms of ${\cal L}^{(3)}_{\pi N}$ in ref.\cite{mojzis} do not modify
the point-like one-pion exchange eq.(11).  

\begin{center}
\SetScale{0.8}
\SetWidth{1.5}
  \begin{picture}(364,72)
\Line(0,0)(0,85)
\Line(57,0)(57,85)
\DashLine(0,42.5)(57,42.5){7}

\Line(85,0)(85,85)
\Line(142,0)(142,85)
\DashLine(85,42.5)(142,42.5){7}
\DashCArc(113.5,56.7)(14.2,0,360){7}

\Line(199,0)(199,85)
\Line(256,0)(256,85)
\DashLine(199,42.5)(256,42.5){7}
\DashCArc(184.8,42.5)(14.2,0,360){7}

\Line(299,0)(299,85)
\Line(356,0)(356,85)
\DashLine(299,42.5)(356,42.5){7}
\DashCArc(299,63.8)(14.2,90,270){7}

\Line(398,0)(398,85)
\Line(455,0)(455,85)
\DashLine(398,42.5)(455,42.5){7}
\DashCArc(398,42.5)(20,90,270){7}
  \end{picture}

\medskip

{\it Fig.1: One-pion exchange, tree-level diagram and diagrams with
one-loop corrections}

\end{center}
\subsection{IRREDUCIBLE TWO-PION EXCHANGE}
The two-pion exchange diagrams are shown in Fig.2. Their contributions to the 
NN T-matrix at second order in small momenta come from insertions of vertices 
and propagators of the leading order $\pi N$-Lagrangian ${\cal L}_{\pi N}^{(1)
}$ alone. All these diagrams can be evaluated in a straightforward manner using
the heavy baryon formalism of ref.\cite{review}, with the exception of the last
one, the planar box graph. Here the problem of the so-called pinch singularity
arises \cite{weinb}. In the heavy mass limit the two nucleon propagators lead
to an expression, $[(l_0+ i \eta)(l_0- i \eta)]^{-1}$ with infinitesimal
 $\eta>0$. The integration contour along the real $l_0$-axis is
pinched between two nearby poles at $\pm i \eta$. A loop integral involving
this product of propagators does not exist since the two poles at infinitesimal
distance $2\eta$ let the result diverge as $1/\eta$. In ref.\cite{weinb} is was
therefore suggested that one should give up covariant perturbation theory and
switch over to "old fashioned" time-ordered perturbation theory in order to
avoid nearly vanishing energy denominators. This strategy was then carried out
in the work of Ordonez, Ray and van Kolck \cite{kolck}. 

\begin{center}
\SetScale{0.8}
\SetWidth{1.5}
  \begin{picture}(364,72)
\Line(0,0)(0,85)
\Line(57,0)(57,85)
\DashCurve{(0,42.5)(28.5,64)(57,42.5)}{7}
\DashCurve{(0,42.5)(28.5,21)(57,42.5)}{7}

\Line(100,0)(100,85)
\Line(157,0)(157,85)
\DashLine(100,42.5)(157,71){7}
\DashLine(100,42.5)(157,14){7}

\Line(199,0)(199,85)
\Line(256,0)(256,85)
\DashLine(199,71)(256,42.5){7}
\DashLine(199,14)(256,42.5){7}

\Line(298,0)(298,85)
\Line(355,0)(355,85)
\DashLine(298,71)(355,14){7}
\DashLine(298,14)(355,71){7}

\Line(397,0)(397,85)
\Line(454,0)(454,85)
\DashArrowLine(454,65)(397,65){7}
\DashArrowLine(397,20)(454,20){7}
\Text(340,56)[b]{$l+q$}
\Text(340,20)[b]{$l$}
\Text(321,0)[lb]{$p_1$}
\Text(360,0)[rb]{$p_2$}

  \end{picture}

\medskip

{\it Fig.2: Two-pion exchange diagrams at one-loop order}

\end{center}

The abovementioned problem is of purely kinematical origin. It comes from 
taking the infinite nucleon mass limit $ M\to \infty$ from the very beginning. 
In a relativistic calculation with finite nucleon mass $M$ the planar box
diagram is well defined, modulo ultra-violet divergences which can be handled
by dimensional regularization.  The planar box diagram includes the iterated
one-pion exchange which is usually  generated by the non-relativistic
Lippmann-Schwinger equation in the form $U^{(1\pi)} {\cal G}_{NN} U^{(1\pi)}$
\cite{erwe}. The non-relativistic two-nucleon propagator ${\cal G}_{NN}$ is the
inverse kinetic energy difference (between initial and intermediate state) and 
thus proportional to the nucleon mass $M$. It is exactly this factor
of $M$ in the iterated one-pion exchange which makes the heavy baryon limit of
the planar box graph divergent and ill-defined. 

The task is to separate off the
iterated one-pion exchange proportional to $M$ and to find the proper 
irreducible part which then exists in the infinite nucleon mass limit, $M\to
\infty$. For that purpose one considers the product of the four relativistic
propagators entering  the planar box diagram and  performs the
$l_0$-integration via contour methods before doing the $1/M$-expansion,
\begin{eqnarray}& & \int {dl_0\over 2\pi } {4 M^2 \,i\over 
 [(p_1-l)^2-M^2+i\eta] [(p_2+l)^2-M^2+i\eta][(l+q)^2-m_\pi^2+i\eta]
[l^2-m_\pi^2+i\eta]} \nonumber\\ & &\qquad \qquad \quad = {M \over [p^2
- (\vec p -\vec l\,)^2 + i\eta]\, \omega_1^2 \omega_2^2} + {\omega_1^2+
\omega_1 \omega_2+\omega_2^2 \over 2 \omega_1^3\omega_2^3 (\omega_1+\omega_2)
} + {\cal O}(M^{-1}) \,. \end{eqnarray} 
Here $p_1^\mu= (E, \vec p\,),\,\, p_2^\mu =(E,-\vec p\,)$ and $q^\mu = (0, \vec
q\,)$ are the initial four-momenta of the nucleons (see Fig.2) and the 
four-momentum transfer in the CM frame, $\omega_1 = \sqrt{\vec l\,^2 +m_\pi^2}$
and $\omega_2 = \sqrt{(\vec l + \vec q\,)^2 +m_\pi^2}$ are the on-shell
energies of the two exchanged pions. The factor $4M^2$  was put into the 
numerator in order to have the correct heavy baryon limit. In the $l_0$-plane 
the integrand in eq.(12) has eight simple poles, four below and four above the
real axis. Two of these poles (one above and one below the real axis) move 
towards each  other for increasing nucleon mass $M$. The integral in eq.(12) 
can be easily evaluated by closing the $l_0$-contour e.g. in the lower 
half-plane and using residue calculus. The first term in the second line of 
eq.(12), the "iterated one-pion exchange", comes just from that pole which is 
moving towards another one. The second term, the "irreducible two-pion 
exchange", arises from the two poles of the pion propagators. The fourth pole 
leads to a strongly suppressed contribution of order $M^{-5}$. The separation 
into "iterated one-pion exchange" and "irreducible two-pion exchange" as 
obtained in eq.(12) is not the standard one of time-ordered  perturbation
theory, which gives $-[2\omega_1 ^2 \omega_2^2 (\omega_1+\omega_2)]^{-1}$ for
the irreducible part. Nevertheless, eq.(12) is the correct result and it 
perfectly agrees with the one of time-ordered perturbation theory when  
performing the systematic $1/M$-expansion. Let $\delta T =[(\vec p-\vec
l\,)^2-p^2)] /2M $ denote the (single) nucleon kinetic energy difference
between  intermediate and initial (or final) state. From the two irreducible
and four reducible time-orderings of the planar box diagram one obtains the
following expression for the energy  denominators, 
\begin{eqnarray}& &  {1\over 4 \omega_1\omega_2} \bigg[ -{2\over (\omega_1+
\omega_2 ) (\omega_1+\delta T)(\omega_2+\delta T)} - {4 \over2\delta T(\omega_1
+\delta T ) (\omega_2+\delta T)} \bigg] \nonumber \\& & = -{1 \over 2 \delta
T\, \omega_1^2 \omega_2^2 } +{ \omega_1^2+\omega_1 \omega_2+\omega_2^2 \over 2
\omega_1^3\omega_2^3(\omega_1+\omega_2)}+{\cal O}(\delta T)\,\,,\end{eqnarray} 
including the phase space density factor $(4\omega_1\omega_2)^{-1}$.  
  
One can apply the previous method to the crossed box diagram as well and finds
as a result (in the $M\to \infty$ limit) the same irreducible part as in
eq.(12) but with opposite sign. In the previous analysis the momentum
dependence of the  $\pi NN$-coupling was not included. However, in a leading
non-relativistic approximation it is of the form $\vec \sigma \cdot \vec l$ and
does therefore not interfere with the $l_0$-integration.

As a result one finds that the isoscalar central amplitude $V_C$ adds up to 
zero for the sum of planar and crossed box graphs. For the isovector central
amplitude $W_C$ there is a further relative minus sign from the isospin 
factors, which are  $3 -2 \vec \tau_1 \cdot \vec \tau_2$ for the planar box and
$3 +2 \vec \tau_1 \cdot \vec \tau_2$ for the crossed box. To the amplitude
$W_C$ the two diagrams contribute with equal sign. In the spin-spin and
tensor channels the situation concerning relative signs is just reverse, due to
a further relative sign between planar and crossed box graph coming from the 
different ordering of $\vec \sigma$-matrices. Altogether, one finds from the
graphs in Fig.2  the following (renormalized) contributions to the NN T-matrix
from irreducible two-pion exchange at second order in small momenta, 
\begin{eqnarray} W_C &=&{1\over384\pi^2 f_\pi^4} \bigg\{
\Big[6m_\pi^2(15g_A^4-6g_A^2-1)+q^2(23g_A^4-10g_A^2-1)\Big] \ln{m_\pi \over
\lambda} \nonumber \\ & &  +4m_\pi^2 (4g_A^4+g_A^2+1) +{q^2\over6}
(5g_A^4+26g_A^2+5)\nonumber\\ & &  +\bigg[4m_\pi^2(5g_A^4-4g_A^2-1)
+q^2(23g_A^4 -10g_A^2-1) +{48g_A^4 m_\pi^4\over 4m_\pi^2+q^2}\bigg] L(q) 
\bigg\}\,,  \\   V_T&=&-{1\over q^2} V_{S}= {3g_A^4 \over 64\pi^2
f_\pi^4}\bigg\{ \ln{m_\pi \over \lambda} -{1\over 2} + L(q) \biggr\} \,, 
 \end{eqnarray}  
with the logarithmic loop function
\begin{equation} L(q)  ={w\over q} \ln {w+q \over 2m_\pi}\,, \qquad w=
\sqrt{4m_\pi^2+q^2} \,. \end{equation}
The NN-amplitudes not specified in eqs.(14,15) are zero. The first two lines in
eq.(14) for $W_C$ and the constant terms in eq.(15) for $V_T$ do not contribute
to phase shifts with $L\geq2$ and mixing angles with $J\geq2$. Such polynomials
in $q^2$ drop out of the projection formulas eqs.(6-9) by  orthogonality of the
$P_J(z)$. Therefore the value of the (a priori arbitrary) renormalization scale
$\lambda$ is irrelevant in this work.

Next we consider the two-pion exchange contributions at third order in the 
small momentum expansion. These arise from the one-loop diagrams in Fig.2 with
exactly one insertion from the next-to-leading order $\pi N$-Lagrangian 
${\cal L}_{\pi N}^{(2)}$. The contributions are either proportional to the 
low-energy constants $c_i$ or they carry a suppression factor $1/M$. Again, all
but the planar box graph can be evaluated in straightforward manner using the
heavy baryon formalism \cite{review}. The planar box graph requires a special 
treatment in order to obtain correctly all $1/M$-corrections. First, the
expansion of the propagator integral, eq.(12), in powers of $1/M$ has to be 
performed one order further. This gives rise to new irreducible pieces
(proportional to $1/M$) and a relativistic correction factor, $1-p^2/2M^2
\simeq M/E$, to the iterated one-pion exchange. Secondly, one has to expand in 
powers of $1/M$ the product of pseudovector $\pi NN$-vertices and ($p\!\!\!/+M
)/2M$-factors from the nucleon propagators sandwiched between in- and out-going
Dirac spinors. At order $1/M$, only terms proportional to $l_0$ and $l_0^3$ are
generated this way. The integral analogous to eq.(12) with an additional factor
$l_0$ or $l_0^3$ in the numerator gives zero, $0+ {\cal O}(M^{-2})$. At order
$1/M^2$ terms independent of $l_0$ but proportional to $\vec l\cdot (2\vec p
-\vec l\,)$ appear. Together with eq.(12) these give rise to a further 
$1/M$-correction to the irreducible two-pion-exchange. With these techniques
and the rules of dimensional 
regularization to evaluate the remaining $d^3 l$-integral one obtains all
$1/M$-corrections coming from the planar box graph. As a check we applied the
same technique to the crossed box and triangle graphs. In all cases we found
perfect agreement with the results obtained independently using the heavy
baryon formalism. This shows that our method of doing first the $l_0$-integral
and then expanding in $1/M$, which is necessary for a systematic evaluation of
the planar box diagram, is indeed correct. Putting all pieces together, we
finally arrive after tedious calculation at the following irreducible two-pion
exchange contributions to the NN T-matrix at third order in small momenta,  
\begin{eqnarray} V_C &=&{3g_A^2 \over 16\pi f_\pi^4} \bigg\{ 4(c_1-c_3)m_\pi^3
-{g_A^2m_\pi \over 16 M}(m_\pi^2+3q^2) - {g_A^2 m_\pi^5  \over 16M(4m_\pi^2 
+q^2)} \nonumber \\ & &  -c_3m_\pi q^2 +\Big[2m_\pi^2( 2c_1-c_3)-q^2  \Big(c_3
+{3g_A^2\over16M}\Big)\Big](2m_\pi^2+q^2) A(q) \biggr\}\,, \end{eqnarray} 
with a new loop function
\begin{equation} A(q)={1\over 2q}\arctan{q \over 2m_\pi}\,, \end{equation}
\begin{eqnarray} W_C &=& {g_A^2\over128\pi M f_\pi^4} \bigg\{ (8-11g_A^2) 
m_\pi^3 +(2-3g_A^2) m_\pi q^2 -{3g_A^2 m_\pi^5\over 4m_\pi^2 +q^2 } \nonumber 
\\ & &  + \Big[ 4m_\pi^2 +2q^2-g_A^2(4m_\pi^2+3q^2) \Big] (2m_\pi^2+q^2) A(q)
\bigg\} \,,\\ V_T &=& -{1
\over q^2} V_{S}=-{9g_A^4\over512\pi Mf_\pi^4} \Big\{ m_\pi+ (2m_\pi^2+q^2) 
A(q) \Big\}  \,,  \\ W_T &=&-{1\over q^2}W_{S} = {g_A^2\over 32\pi f_\pi^4}
\bigg\{ \Big( c_4 +{2-3g_A^2 \over 8M} \Big) m_\pi \nonumber \\ & &  + \bigg[
\Big( c_4 +{1\over 4M} \Big)
(4m_\pi^2+q^2)-{g_A^2 \over 8M} (10m_\pi^2+3q^2)  \bigg] A(q) \bigg\}\,,\\
V_{SO} &=&  {3g_A^4  \over 64\pi M f_\pi^4}\Big\{ m_\pi + 
(2m_\pi^2+q^2) A(q) \Big\}  \,,\\  W_{SO} &=& {g_A^2(1-g_A^2)\over 64\pi M 
f_\pi^4} \Big\{ m_\pi +  (4m_\pi^2+q^2) A(q) \Big\}\,. \end{eqnarray}   
Note that the $c_1$-part of $V_C$ in eq.(17) is proportional to the
one-loop contribution to the nucleon scalar form factor, namely $-8c_1
f_\pi^{-2}\,\sigma_N(-q^2)_{loop}$ \cite{form} and the low-energy constant 
$c_3$ is related to the so-called nucleon axial polarizability
\cite{review,ericson}, $\alpha_A = -2c_3/f_\pi^2-g_A^2m_\pi(48g_A^2+77)/(384\pi
f_\pi^4)= 6.44$  fm$^3$ which is governed by virtual $\Delta$-isobar 
excitation. Similarly, the $c_4$-part of $W_T$ in eq.(21) is proportional to
the one-loop contribution to the nucleon isovector magnetic form factor
(normalized to $1+\kappa_p-\kappa_n$), namely $-(c_4/4M f_\pi^2) \, G_M^V(-q^2)
_{loop}$. The low-energy constant $c_2$ is absent in the above expressions. In
the actual calculation one sees that this comes from the fact that the energy
transfer in the CM frame is zero, $q_0=0$.  We note that all contributions at 
third order in small momenta are finite (free of divergences). This is 
consistent with the fact that no local counter terms (analytic in the quark 
mass) exist at this order of the small momentum expansion to cancel
divergences. The spin-orbit amplitudes $V_{SO},\, W_{SO}$ generated by chiral 
two-pion exchange are proportional to $1/M$. This is different from the 
OBE-model, where scalar and vector meson exchange lead to spin-orbit terms
proportional to $1/M^2$. Furthermore, to the order we are working, there is no
quadratic spin-orbit contribution from irreducible graphs. Such terms are 
suppressed by a factor $1/M^2$. Already a short glance at the analytical 
formulas, eqs.(14-23) for the two-pion exchange contributions shows that
these are all increasing (in magnitude) with increasing momentum transfer $q$. 
This is in contrast to the OBE-model, where by construction all NN-potentials
decrease  as a function of $q$. The growth in $q$ of the chiral two-pion
exchange potentials is a consequence of the Goldstone boson nature of the pion.
A Goldstone boson has (and induces) weak interaction at low energies and
stronger interaction at higher energies (see also the explicit expressions for
the one-loop $\pi N$-scattering amplitude in ref.\cite{pin} and the one-loop
$\pi\pi$-scattering amplitude in ref.\cite{leut}, which share the same 
property). However, in many model calculations, this feature is altered by the
introduction of phenomenological cut-offs, form factors and the like.  

\subsection{ITERATED ONE-PION EXCHANGE}
The planar box graph includes the iterated one-pion exchange, which is enhanced
by a factor $M$. Even though we counted the diagram naively as second
order in small momenta, the actual contribution to the NN T-matrix is of first
order in small momenta. The occurrence of such a large scale enhancement factor
$M$ in iterated diagrams changes the correspondence between the loop expansion
and the small momentum expansion for processes involving two (or more)
nucleons. Ladder diagrams with $n$ loops contribute already at $n$-th order in
the small momentum expansion, whereas the naive counting rules would classify
them as order $2n$. The iterated one-pion exchange including the relativistic 
correction factor $M/E$ has the following integral  representation,
\begin{equation}{g_A^4M^2 \over 16E f_\pi^4} (2\vec \tau_1\cdot \vec \tau_2 -3)
\int {d^3 l \over (2\pi)^3} {\vec \sigma_1\cdot (\vec l +\vec p\,')\, \vec
\sigma_2\cdot (\vec l +\vec p\,')\, \vec \sigma_1\cdot (\vec l +\vec p\,)\,
\vec  \sigma_2\cdot (\vec l +\vec p\,) \over (p^2 - l^2+i \eta) [(\vec l+\vec
p\,')^2 +m_\pi^2]  [(\vec l+ \vec p\,)^2 +m_\pi^2] }\,. \end{equation}
The isospin factor $8I-9$ suppresses the contribution to $(I=1)$-states by a
factor $1/9$ compared to $(I=0)$-states. The resulting three-dimensional
integrals can be evaluated in closed form, and the contributions to the NN
T-matrix are expressed in terms of the following set of  complex-valued
functions, 
\begin{eqnarray} \Gamma_0(p) &=& {1\over 2p} \bigg[  \arctan{2p\over m_\pi} +
i\, \ln{u\over m_\pi} \bigg]\,, \quad u = \sqrt{m_\pi^2+4p^2}\,, \\ 
\Gamma_1(p) &=& {1\over 2p^2} \Big[ m_\pi + i\, p -(m_\pi^2+2p^2) \Gamma_0(p) 
\Big] \,,\\  G_0(p,q) &=& {1\over q \,R} \bigg[  \arcsin{q \,m_\pi\over u\,w}
+ i\, \ln {p\,q+R\over u\,m_\pi } \bigg]\,,  \quad R = \sqrt{m_\pi^4 + p^2
w^2}\,, \\ G_1(p,q) &=& { \Gamma_0(p) -2 A(q) -(m_\pi^2 +2p^2) G_0(p,q)
\over 4p^2-q^2} \,,\\ G_2(p,q) &=& p^2 G_0(p,q) +(m_\pi^2 +2p^2) G_1(p,q) +A(q)
\,, \\ G_3(p,q) &=& {{1\over 2} \Gamma_1(p) -p^2 G_0(p,q) - 2(m_\pi^2
+2p^2) G_1(p,q) \over 4p^2-q^2} \,,\end{eqnarray}
with $w$ and $A(q)$ given in eqs.(16,18). Note that the functions 
$G_{1,2,3}(p,q)$ are not singular at $z=-1$ or $q^2=4p^2$.  In terms of these
complex functions the contributions to the NN T-matrix read, 
\begin{eqnarray} W_C&=&-{2\over 3} V_C =  {g_A^4 M^2 \over 128 \pi E f_\pi^4}
\Big\{4(2m_\pi^2+q^2) \Gamma_0(p)+2 q^2\Gamma_1(p) -4i\,p - (2m_\pi^2+q^2)^2
\,G_0(p,q) \Big\} \,, \nonumber \\ & & \\ W_T &=& -{2\over 3} V_T= -{1\over
q^2} W_{S} ={2 \over 3q^2} V_{S}=-{g_A^4 M^2\over32\pi Ef_\pi^4} \,G_2(p,q)
\,,\\ W_{SO} &=& -{2\over 3} V_{SO} =  {g_A^4M^2\over 64 \pi Ef_\pi^4} \Big\{ 
-2 \Gamma_0(p)-2 \Gamma_1(p)+ (2m_\pi^2+q^2) \big[ G_0(p,q)+2G_1(p,q) \big] 
\Big\}\,, \\ W_Q &=& -{2\over 3} V_Q =  {g_A^4 M^2 \over 32 \pi E
f_\pi^4} \Big\{ G_0(p,q) +4 G_1(p,q)+4 G_3(p,q) \Big\} \,.\end{eqnarray}
Here, for the first time a non-vanishing quadratic spin-orbit contribution is
found. Using dimensional regularization, we find that the iterated one-pion 
exchange is finite and has no divergences. Note that the contributions of
iterated one-pion exchange, eqs.(31-34), depend on both kinematical variables
$p$ and $q$, whereas irreducible two-pion exchange, eqs.(14-23), depends only 
on the momentum transfer $q$. To the order we are working the iterated one-pion
exchange is the only contribution to the NN T-matrix with a non-vanishing
imaginary part and thus it restores unitarity to this order. In partial waves 
this imaginary part is given by the square of the phase-shift calculated from
one-pion exchange, eq.(11); in the case of the mixing triplet states with $L=
J \pm 1$ it is actually a quadratic expression in these phase shifts and the
mixing angle. Therefore the imaginary part of iterated one-pion exchange allows
for a numerical check of unitarity, which actually serves as a check of both,
the analytical expressions for iterated one-pion exchange, eqs.(31-34) and the
projection formulas, eqs.(6-9). We have found that in all cases this numerical
check of unitarity works  at an accuracy $10^{-4}$ or better. Furthermore, we
have verified analytically that the optical theorem is satisfied. It equates
the imaginary part of the NN T-matrix in forward direction (only $V_C$ and
$W_C$ in eq.(31) contribute) to $pE/M^2$ times the (unpolarized) total NN-cross
section  calculated in (non-relativistic) one-pion exchange approximation
without crossed nucleon lines.  
     
Since each iteration of one-pion exchange brings a further enhancement factor
$M$, up to three iterations will contribute to the NN T-matrix calculated
completely to third order in small momenta. This corresponds to two- and
three-loop diagrams which are extremely difficult to evaluate. Numerically, 
we find that in most cases iterated one-pion exchange is small compared to 
irreducible two-pion exchange. We can therefore assume that the higher order 
iterations can be neglected.
  
At zeroth order in the small momentum expansion, not only one-pion exchange
eq.(11) contributes to the NN T-matrix, but also contact terms of the
(Fierz-antisymmetric) form,
\begin{equation} 3B+ B'\, \vec \tau_1 \cdot \vec \tau_2 - (2B+B')\, \vec 
\sigma_1 \cdot \vec \sigma_2- B\, \vec \tau_1 \cdot \vec \tau_2 \, \vec 
\sigma_1 \cdot \vec \sigma_2 \,. \end{equation} 
In tree approximation eq.(35) gives a contribution to the two S-wave amplitudes
only. At one-loop order the iteration of this contact interaction and diagrams
with an  additional pion exchange occur. We have checked that the one-loop
diagrams involving the lowest order contact interaction eq.(35) do not
contribute to the phase-shifts  with $L\geq 2$ and the mixing angles with
$J\geq2$.  Therefore, we do not need  to specify the constants $B,\,B'$ in
eq.(35).  

This completes the discussion of the NN T-matrix in one-loop approximation. 
We find that if we restrict ourselves to the phase-shifts with $L\geq2$ and
mixing angles with $J\geq2$ (peripheral partial waves), not even a single 
adjustable parameter shows up. To the order we are working this peripheral part
of the NN T-matrix involves only well-known physical parameters such as $g_{\pi
N}, \, f_\pi,\, m_\pi,\, M$ and three low-energy constants $c_{1,3,4}$ which
are well determined from $\pi N$-scattering data. Such features  allow for a 
clean test of chiral symmetry in NN-interaction.

\section{RESULTS AND DISCUSSION}
In this section we present and discuss our results for the phase-shifts with
$L\geq 2$ and mixing angles with $J\geq 2$ up to nucleon laboratory kinetic
energies of $T_{lab} = 280$ MeV, the $NN\pi^0$-threshold.  At such an energy, 
momentum transfers up to $q= 5.25 \, m_\pi =725$ MeV are involved, which are 
quite large for the application of chiral perturbation theory. Despite this 
fact, we present here results up to the $NN\pi^0$-threshold, in order to 
demonstrate where our model independent predictions are in agreement with
existing data and where deviations show up. For the sake of completeness, we
summarize again the values  of the physical parameters used in the calculation:
$f_\pi = 92.4$ MeV, $g_{\pi N}=13.4$ as determined from $\pi N$-dispersion
analysis \cite{hoehler}, $m_\pi = 138$ MeV, $M=939$ MeV for the average pion
and nucleon mass and the central values of the low-energy constants, $c_1=-0.9$
GeV$^{-1}$, $c_3 = -5.3$ GeV$^{-1}$ and $c_4=3.6$ GeV$^{-1}$ \cite{pin}.

\subsection{D-WAVES} 

The D-wave phase shifts and mixing angle $\epsilon_2$ are shown in Fig.3. The
dashed curve corresponds to the one-pion exchange approximation and the full
curve includes in addition two-pion exchanges (irreducible two-pion exchange
and iterated one-pion exchange). The dotted and dashed-dotted curves represent
the empirical NN phase shift analyses of ref.\cite{arnd} and ref.\cite{swart}
(as far as available), respectively. In all cases the model independent 
two-pion exchange corrections go into the proper direction, but they tend to be
too large, already at small $T_{lab}$. The one-loop predictions for the D-wave
phase shifts deviate appreciably from the data. An exception is the mixing
angle $\epsilon_2$ which 
is well reproduced up to $T_{lab} = 120$ MeV. The $2\pi$-exchange tensor force
in this channel has the correct sign. However, it grows too strongly with
energy and even overcompensates the $1\pi$-exchange tensor force at the 
$NN\pi^0$-threshold. The $^3D_1$ phase shift is also in fair agreement with
the data up to $T_{lab} = 200$ MeV. This results from an almost complete 
cancelation of irreducible two-pion exchange and iterated one-pion exchange
contributions.  Above $T_{lab} =200$ MeV this cancelation mechanism does not
work anymore. The good agreement between the data and the one-pion exchange
approximation in the $^3D_1$ partial wave is contrary to expectations and
appears accidental, since in G-waves (see later) there are still sizeable
differences between data and one-pion exchange approximation. From all this we
conclude that one-pion and two-pion exchange alone are not sufficient to
describe the dynamics in the NN D-waves \cite{bonn} (see also the discussion of
the coordinate potentials in the next section). Obviously, strong cancelations
between $2\pi$-exchange and $3\pi$-exchange together with shorter range effects
are taking place here. The latter ingredients of the NN T-matrix go beyond the
one-loop approximation.  

\subsection{F-WAVES}   
The F-wave phase shifts and the mixing angle $\epsilon_3$ are shown in Fig.4.
Again, the $2\pi$-exchange corrections to  $1\pi$-exchange go into the right
direction. The predictions for the phase-shifts in the $^1F_3$ and
$^3F_3$ partial waves are in good agreement with existing data up to
$T_{lab}=180$ MeV. At larger energies the $2\pi$-exchange effects become again
too strong. In the $^3F_2$ partial wave we find a correction to $1\pi$-exchange
opposite to the trend of the data of ref.\cite{swart}. The latest fit of the
$^3F_2$ phase shift from the VPI group \cite{arnd} (not shown) is very similar
to the one of ref.\cite{swart}. The mixing angle  $\epsilon_3$, however, is in
perfect agreement with the data of ref.\cite{arnd,swart} for all energies up to
the $NN\pi^0$-threshold at $T_{lab}=280$ MeV. To this observable irreducible
$2\pi$-exchange and iterated $1\pi$-exchange contribute with almost equal
strength. Note that the empirical F-wave phase-shifts are quite small, namely
less than $4.5^\circ$. It would be very interesting to see how the present
discrepancies between the chiral NN F-wave phase shifts and the existing
analyses reflect themselves in a direct comparison with  scattering data,
i.e. differential cross sections and polarization observables.  Unfortunately,
we can not present such a direct comparison here. 

\subsection{G-WAVES}   
The G-wave phase shifts and the mixing angle $\epsilon_4$ are shown in Fig.5. 
The chiral predictions are in good agreement with data for all four partial
wave phase shifts and the mixing angle up to $T_{lab}=220$ MeV. Note that the 
differences between $1\pi$-exchange and data are still sizeable in the $^1G_4$
and $^3G_5$ partial waves. The model independent chiral $2\pi$-exchange 
corrections  have indeed the correct sign and magnitude to close this gap 
between data and $1\pi$-exchange approximation.  In the $^1G_4$ partial wave
the correction comes mainly from irreducible $2\pi$-exchange, whereas in the
$^3G_5$ partial wave (with $I=0$) both irreducible $2\pi$-exchange and iterated
$1\pi$-exchange contribute with roughly equal strength. For the $^3G_3$ and
$^3G_4$ phase shifts and $\epsilon_4$ the $2\pi$-exchange corrections are quite
small. Nevertheless, these small effects improve the agreement between data and
chiral predictions.

\subsection{H-WAVES}
The H-wave phase shifts and the mixing angle $\epsilon_5$ are shown in Fig.6. 
The corrections due to $2\pi$-exchange are already quite small and they follow
the trend of the data. An exception is the $^3H_4$ phase shift where we find a
very small positive contribution. However, the empirical $^3H_4$ phase shifts
are very small, less than $0.5^\circ$, and the difference between the existing
phase shift analyses and the chiral prediction may be insignificant if one
compares directly the NN-scattering data. 

\subsection{I-WAVES}
The I-wave phase shifts and the mixing angle $\epsilon_6$ are shown in Fig.7.
Again, the corrections coming from $2\pi$-exchange are quite small and they
follow the trend of the data. Altogether, we find a faster convergence to the
$1\pi$-exchange approximation in the high angular momentum partial waves  as
obtained in the existing NN phase-shift analyses. The calculation in chiral 
perturbation theory presented here is most reliable in the high angular 
momentum partial waves. These are the ones which probe the long and medium 
range parts of the NN-interaction which are built in here in a model
independent way via $1\pi$- and chiral $2\pi$-exchange. In the existing
phase shift analysis the (small) effects in high angular momentum partial waves
may result mainly from effective potentials \cite{arnd,swart} fitted to lower
partial waves  with only limited direct constraints from the NN-scattering
data.

\begin{figure}[htbp]
\centerline{\psfig{file=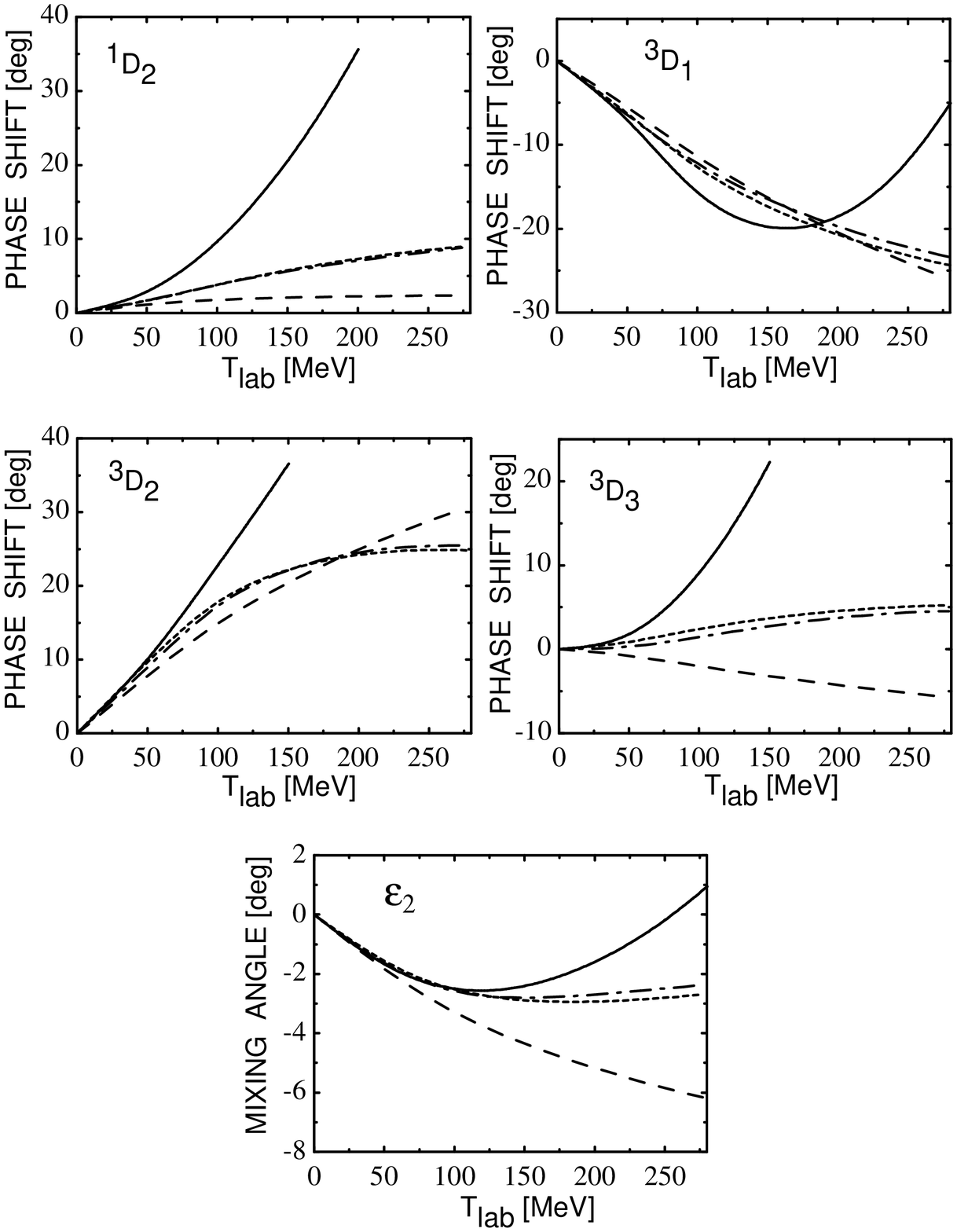,width=14cm}}
{\it Fig.3: D-wave  NN phase shifts and mixing angle $\epsilon_2$ versus 
the nucleon laboratory kinetic energy $T_{lab}$. The dashed curve corresponds
to the $1\pi$-exchange approximation and full curve includes chiral
$2\pi$-exchange as well. The dotted and dashed-dotted curves represent the
empirical NN phase shift analyses of ref.\cite{arnd} and ref.\cite{swart}, 
respectively.}
\end{figure}

\begin{figure}[htbp]
\centerline{\psfig{file=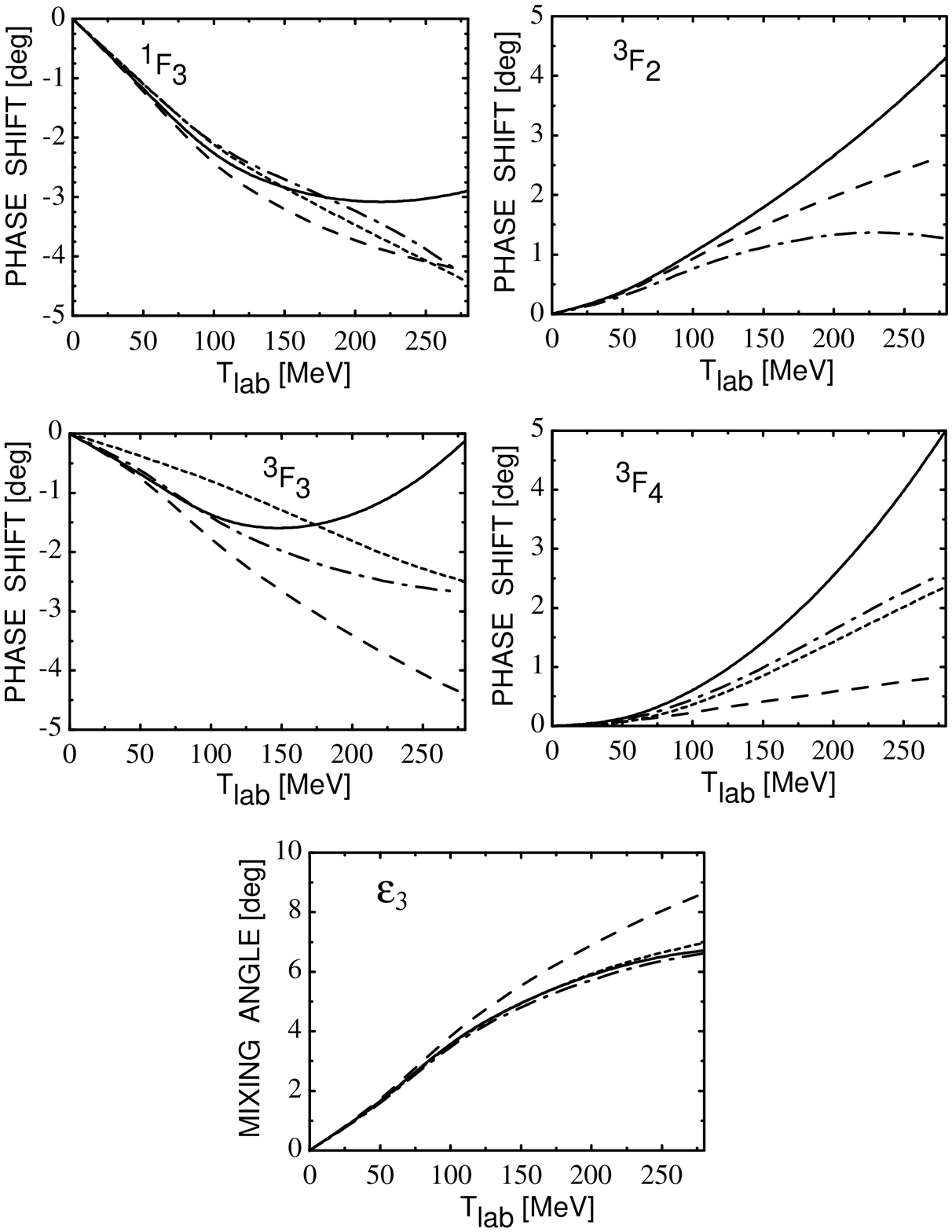,width=15cm}}
{\it Fig.4: F-wave  NN phase shifts and mixing angle $\epsilon_3$ versus the
nucleon laboratory kinetic energy $T_{lab}$. For notation see Fig.3.}
\end{figure}

\begin{figure}[htbp]
\centerline{\psfig{file=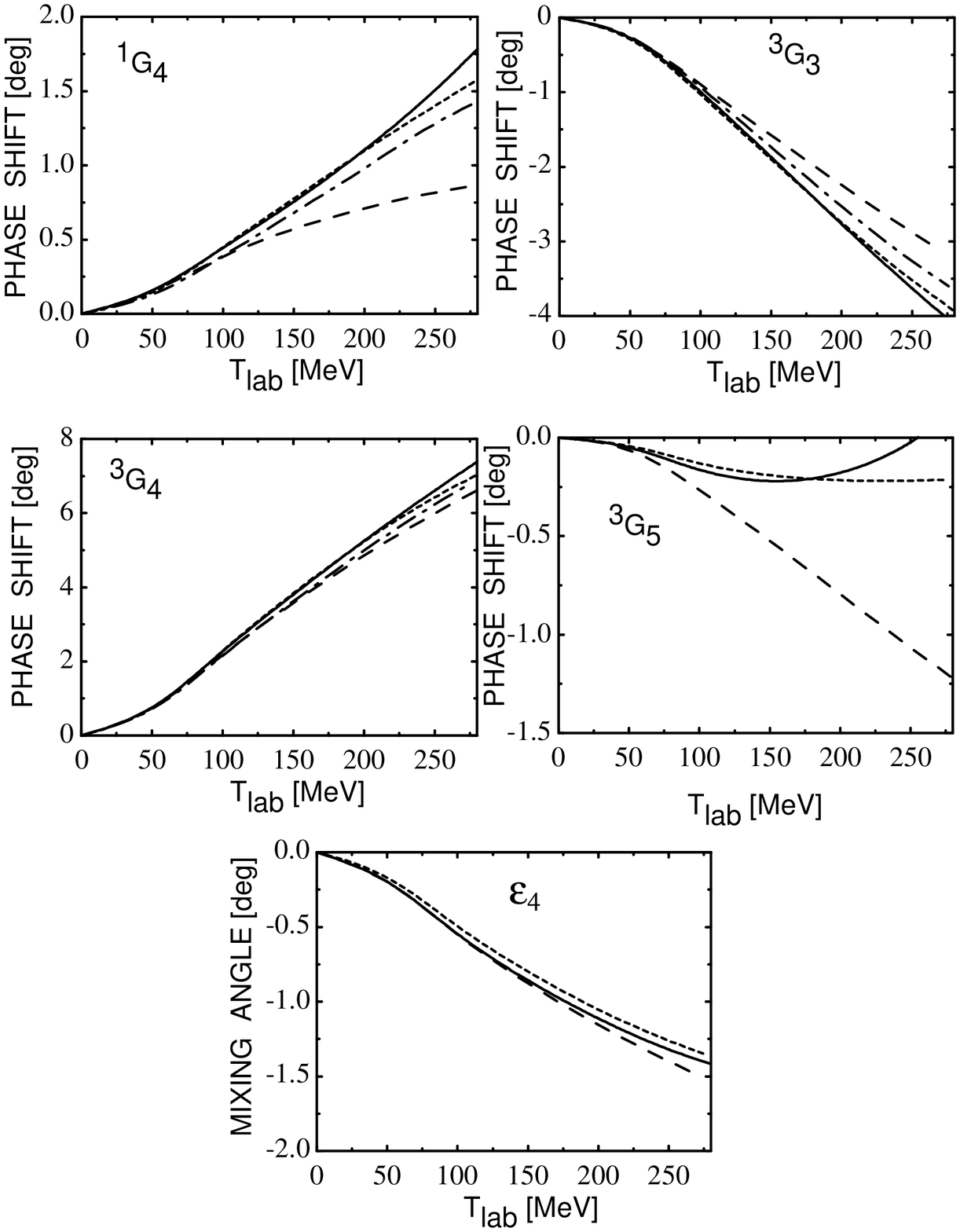,width=15cm}}
{\it Fig.5: G-wave  NN phase shifts and mixing angle $\epsilon_4$ versus the
nucleon laboratory kinetic energy $T_{lab}$. For notation see Fig.3.}
\end{figure}

\begin{figure}[htbp]
\centerline{\psfig{file=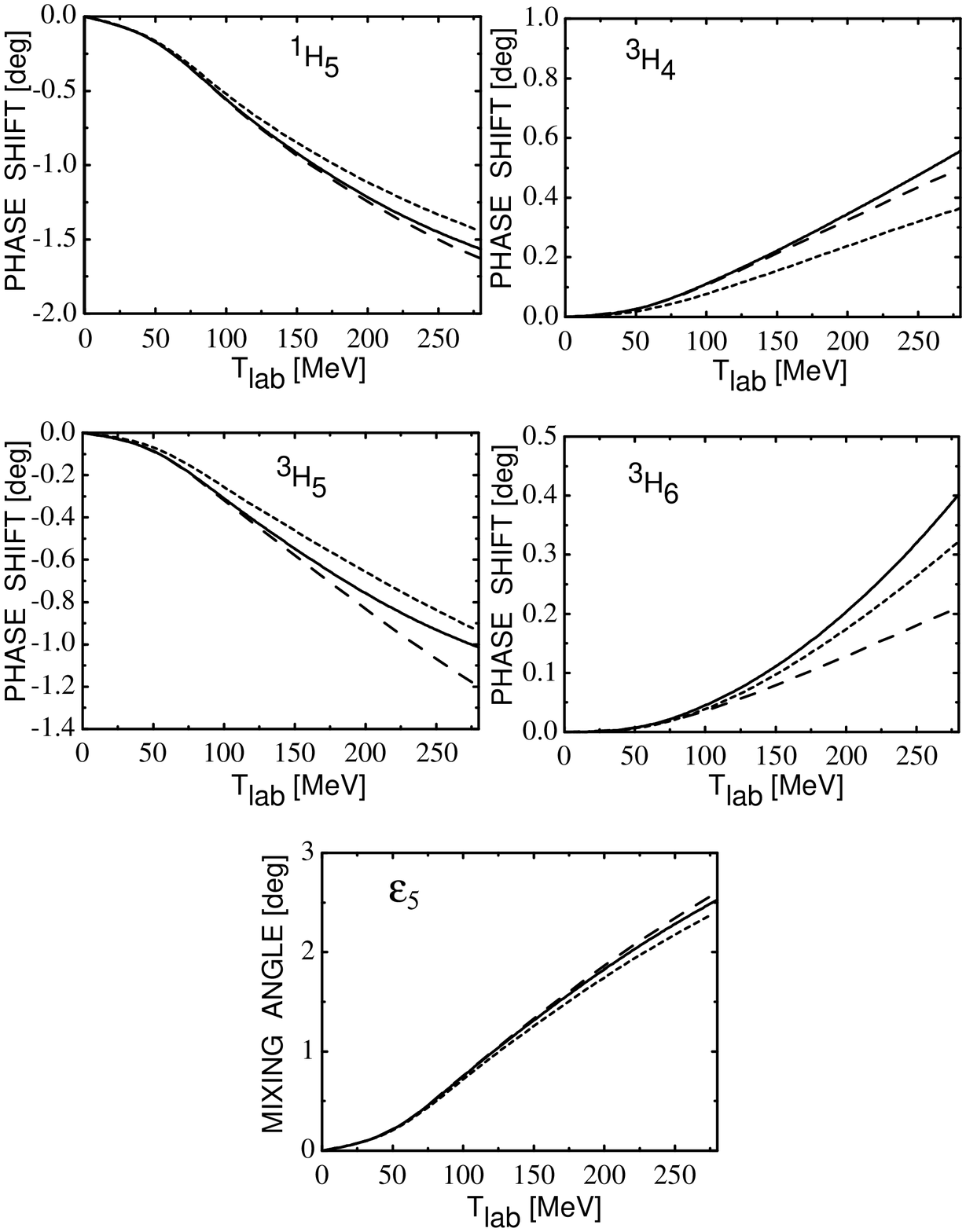,width=15cm}}
{\it Fig.6: H-wave  NN phase shifts and mixing angle $\epsilon_5$ versus the
nucleon laboratory kinetic energy $T_{lab}$. For notation see Fig.3.}
\end{figure}

\begin{figure}[htbp]
\centerline{\psfig{file=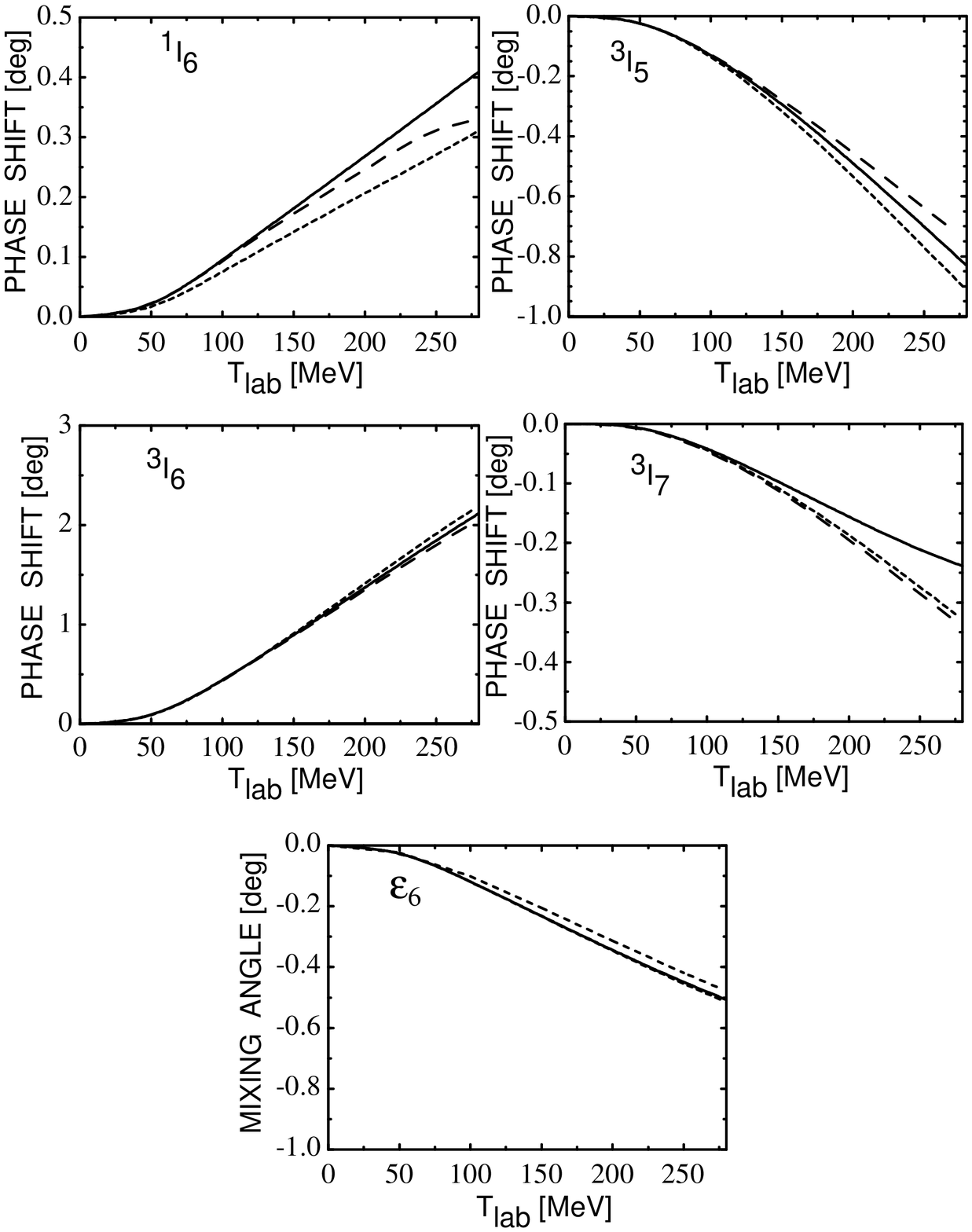,width=15cm}}
{\it Fig.7: I-wave  NN phase shifts and mixing angle $\epsilon_6$ versus the
nucleon laboratory kinetic energy $T_{lab}$. For notation see Fig.3.}
\end{figure}

\section{COORDINATE SPACE REPRESENTATIONS }

In order to interpret the effects of irreducible chiral two-pion exchange 
and to compare with phenomenological approaches  it is desirable to have
a coordinate space representation of the corresponding potentials. However
an ordinary inverse Fourier-transform of the momentum space functions given in
section IIB is not possible, because of their growth with $q$. In this
case the coordinate space representation of the finite range parts can be  
obtained in the form of a continuous superposition of Yukawa functions
\cite{erwe}. The mass spectrum entering in this representation is given by the
imaginary part of the momentum space functions analytically continued to
time-like momentum transfer, $q=0^+-i\,\mu$. For the isoscalar central 
potential in $r$-space one has then the following spectral representation,
\begin{equation} \widetilde V_C(r) = -{1\over 2\pi^2 r} \int_{2m_\pi}^\infty d
\mu \, \mu \, e^{-\mu r}\, {\rm Im}\, V_C(-i \mu) \,\,. \end{equation} 
Using Im\,$A(-i\mu)= (\pi /4\mu) \theta(\mu-2m_\pi)$ for the relevant loop
function (note that due to the coalescence of normal threshold $\mu_0 =2m_\pi$ 
and anomalous threshold $\mu_c = \sqrt{4m_\pi^2-m_\pi^4/M^2} \\=1.995 m_\pi$ 
in the heavy nucleon limit, this imaginary part starts with a non-zero value at
threshold) one finds the following closed expression for the isoscalar
central potential,
\begin{eqnarray} \widetilde V_C(r)&=&{3g_A^2\over32 \pi^2 f_\pi^4} {e^{-2x}
\over r^6} \bigg\{\Big( 2c_1+{3g_A^2\over 16M} \Big) x^2(1+x)^2 +{g_A^2
x^5\over 32M} \nonumber \\ & &+\Big(c_3 +{3g_A^2\over 16M}\Big) (6+12x+10
x^2+4x^3 +x^4)\bigg\} \end{eqnarray} 
with the abbreviation $x=m_\pi r$. Since $c_{1,3} +3g_A^2/16M<0$, this
potential is attractive for all distances $r$ as expected.  Its asymptotic 
behaviour for large $r$ is a modified Yukawa function with
twice the pion mass, $e^{-2m_\pi r}/r^2$, neglecting the small $1/M$-terms.
At a distance $r=1/m_\pi = 1.43$ fm one finds an attraction of $\widetilde
V_C(1.43 {\rm fm}) = - 35.9$ MeV.  This number is twice as large as the
attraction produced by the fictitious $\sigma(550)$-boson of the Bonn potential
\cite{erwe,bonn}. 

\bigskip

\bigskip

\begin{minipage}{7.5cm}
\bild{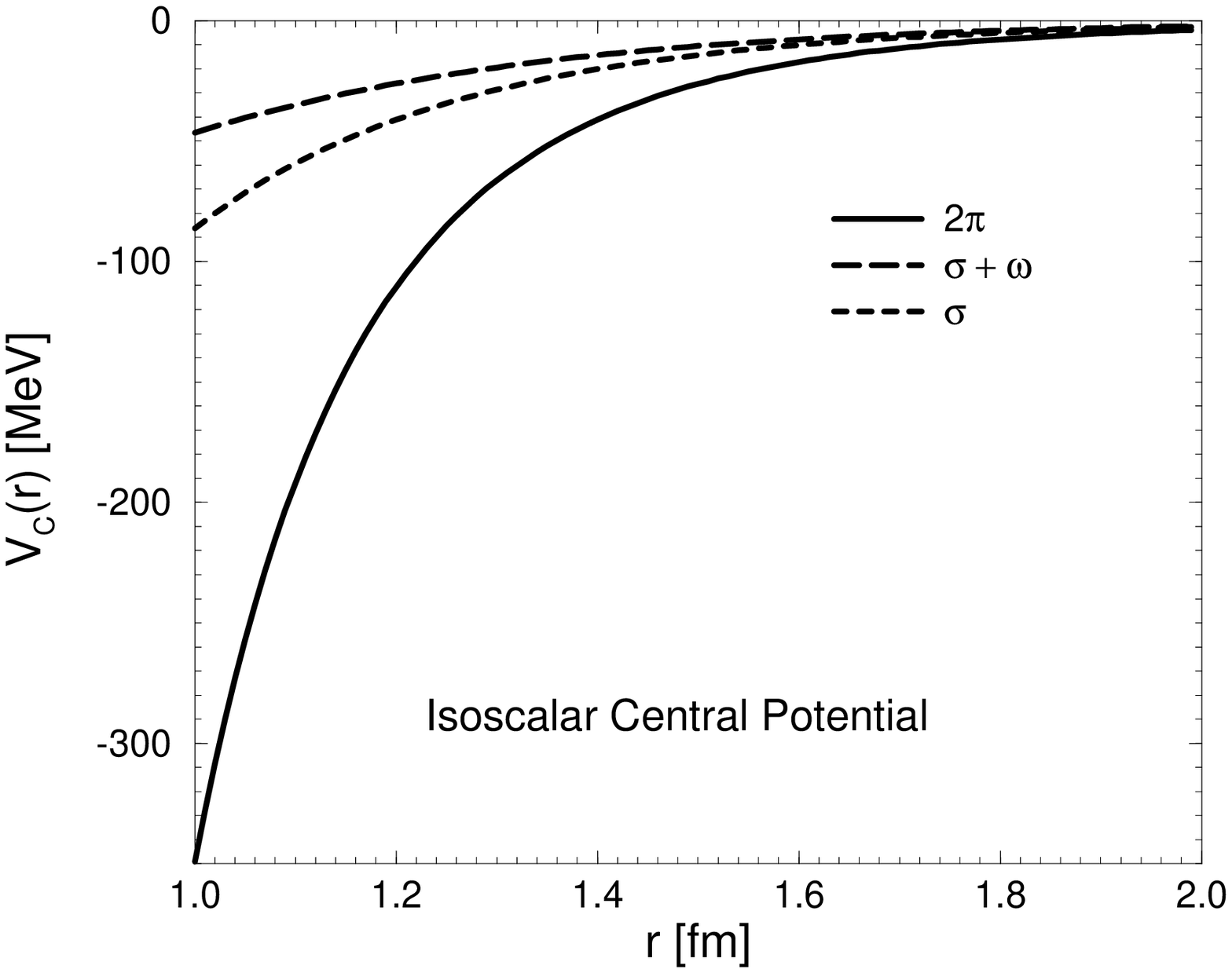}{7.4}
\end{minipage}
\begin{minipage}{7.5cm}
\bild{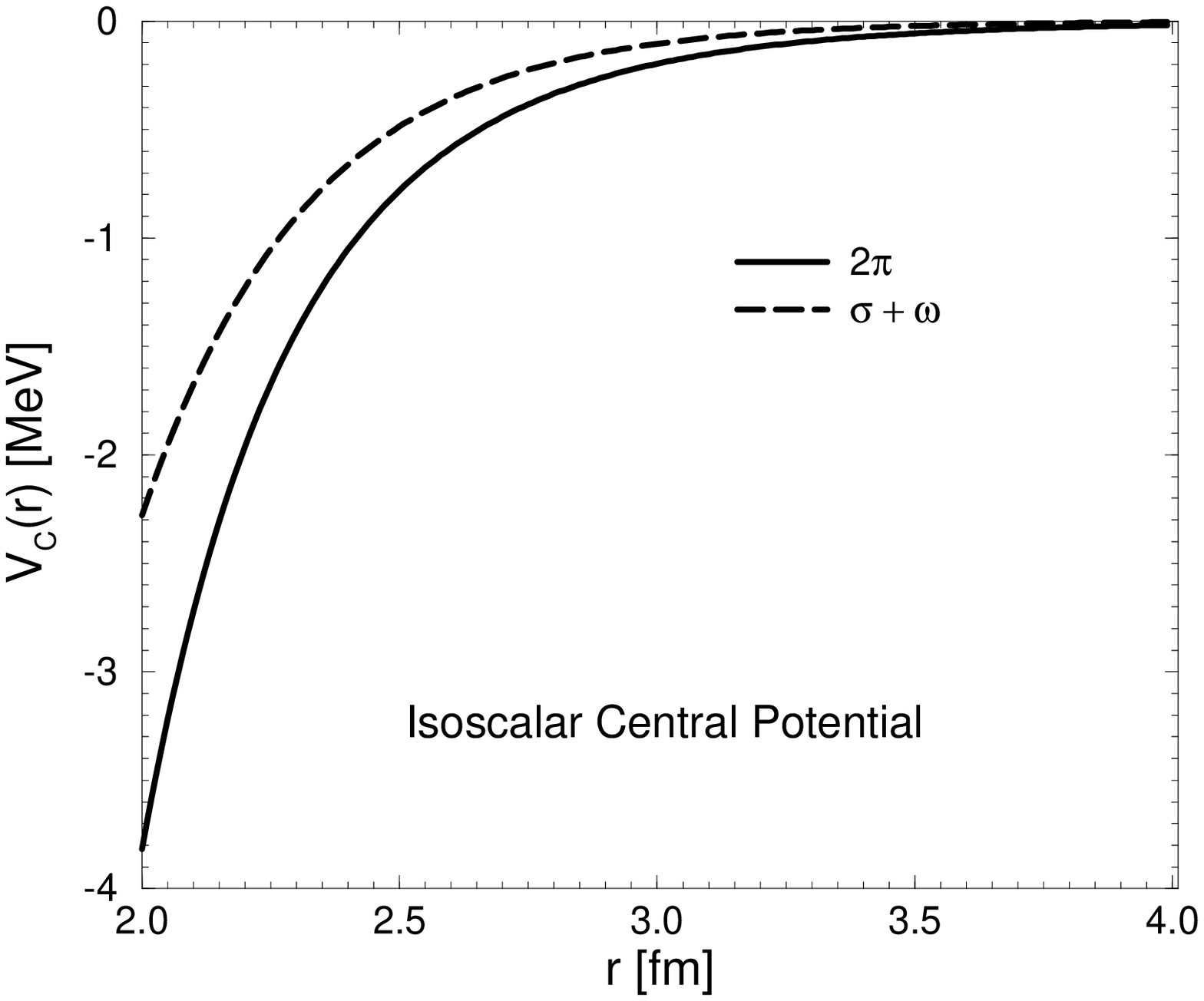}{7.4}
\end{minipage}
{\it Fig.8: The isoscalar central potential $\widetilde V_C(r)$ generated by 
irreducible chiral two-pion exchange in $r$-space (full line). The dashed lines
show the phenomenological $\sigma+\omega$ and $\sigma$-contributions.}

\bigskip

In Fig.8 we show the isoscalar central potential $\widetilde
V_C(r)$ for 1 fm $<r<$ 2 fm and 2 fm $<r<$ 4 fm. Our results are qualitatively
in agreement with the recent work of ref.\cite{robil}, but their isoscalar
central NN-potential is even somewhat more attractive.  The terms proportional
to $c_{1,3}$ in eq.(37) agree with the earlier work of \cite{tarr} up to the
isospin factor $3$ which was inadvertently omitted in \cite{tarr}. Furthermore
it is interesting to observe that in the chiral limit $m_\pi=0$ the isoscalar
central two-pion exchange potential shows a $1/r^6$ fall-off which is typical
for a non-relativistic Van-der-Waals potential. 

The spectral representation of the isovector tensor potential (accompanying the
standard tensor operator $3 \vec \sigma_1 \cdot \hat r \, \vec \sigma_2 \cdot 
\hat r - \vec \sigma_1 \cdot \vec \sigma_2$) takes a similar form, 
\begin{equation} \widetilde W_T(r) = {1\over 6\pi^2 r^3} \int_{2m_\pi}^\infty d
\mu \, \mu \, e^{-\mu r} (3+3 \mu r+ \mu^2r^2)\,{\rm Im} \, W_T(-i \mu) \,\,. 
\end{equation}
The two-pion exchange isovector tensor potential reads explicitly,
\begin{eqnarray}\widetilde W_T^{(2\pi)}(r)&=&{g_A^2\over 48 \pi^2 f_\pi^4} 
{e^{-2x} \over r^6}\bigg\{-\Big(c_4+{1\over 4M}\Big)(1+x)(3+3x +x^2) \nonumber
\\ & & +{g_A^2\over 32M} (36+72x+52x^2+17x^3+2x^4)\bigg\} \,\,.\end{eqnarray}
Since $c_4>0$ it is attractive and therefore cuts down the repulsive isovector
tensor potential due to one-pion exchange
\begin{equation} \widetilde W_T^{(1\pi)}(r) = {g_{\pi N}^2 \over 48 \pi M^2} 
{e^{-x} \over r^3} (3+3x +x^2) \,\,. \end{equation}
At a distance $r=1/m_\pi =1.43$ fm the two-pion exchange isovector tensor
potential is $-22.5\%$ of the one-pion exchange isovector tensor potential. 
Such a reduction is indeed required by phenomenology \cite{erwe}. In Fig.9 we
show the reduced isovector tensor potentials $(m_\pi r)^3 \widetilde W_T(r)$
for distances 1 fm $<r<$ 4 fm. Note that already at $r=2.5$ fm the two-pion
exchange effects start to become visible. The $1/M$-terms in eq.(39) are very
small. 
\bild{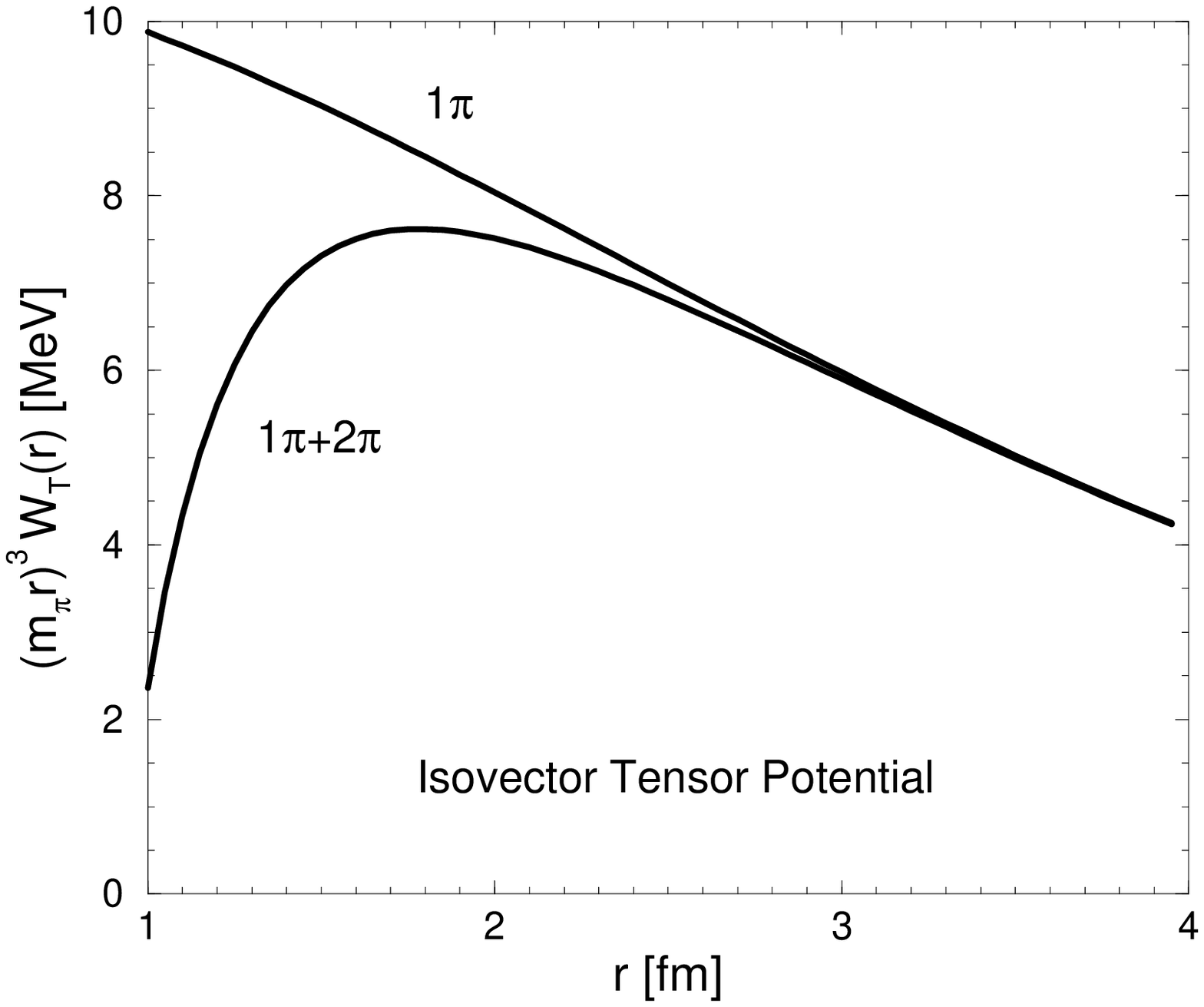}{10}
{\it Fig.9: The reduced isovector tensor potentials $(m_\pi r)^3 \,\widetilde
W_T(r)$  in $r$-space.}

\bigskip

Furthermore, we give the two-pion exchange isoscalar tensor and isovector
central potentials in coordinate space. Using Im\,$L(-i\,\mu) = -(\pi/2\mu)
\sqrt{\mu^2-4m_\pi^2}$ for the relevant loop function, one finds after some
calculation that these can be expressed in terms of two modified 
Bessel-functions,
\begin{eqnarray}\widetilde V_T(r) &=& {g_A^4 m_\pi\over128\pi^3 f_\pi^4\,r^4} 
\bigg\{-12 x\, K_0(2x) -(15+4x^2)\, K_1(2x)\nonumber\\ & & +{3\pi e^{-2x}\over 
8Mr} (12x^{-1}+24+20x+9x^2 +2x^3)  \bigg\} \,\,,
\\ \widetilde W_C(r) &=& {m_\pi \over 128 \pi^3 f_\pi^4\,r^4} \bigg\{\big[1+2
g_A^2(5+2x^2) -g_A^4 (23+12 x^2) \big]  \, K_1(2x) \nonumber \\ & & +
x \big[1+10 g_A^2 -g_A^4(23+4x^2)\big] \, K_0(2x) +{g_A^2 \pi 
e^{-2x}\over 4Mr} \Big[2(3g_A^2-2)\nonumber \\ & & \cdot (6x^{-1}+12+10x+
4x^2+x^3)+g_A^2x(2+4x+2x^2+3x^3) \Big]\bigg\} \,\,.  \end{eqnarray}
The asymptotic behaviour for large distances $r$ is
$e^{-2m_\pi r}/r^{5/2}$ for $\widetilde V_T(r)$ and $e^{-2m_\pi r}/r^{3/2}$ for
$\widetilde W_C(r)$, neglecting the $1/M$-terms. This long range behaviour is
determined by the box graphs (proportional to $g_A^4$). In the chiral limit
$m_\pi=0$ both display a $1/r^5$ powerlike fall-off. Since the modified
Bessel-functions $K_{0,1}(2m_\pi r)$ are positive and $g_A>1$ one sees that
$\widetilde V_T(r)$ and $\widetilde W_C(r)$ are both negative throughout. The
finite range part of the two-pion exchange isovector central potential is
attractive and not repulsive as one would expect from analogies with the
$\rho$-meson. It is very interesting to see how this unexpected attraction
comes about. In the bubble and triangle diagrams (proportional to $g_A^0$ and
$g_A^2$) of Fig.2 a two-pion state with the quantum numbers of the $\rho$-meson
is emitted from one nucleon. As can be seen from the explicit expression in
eq.(42) these processes indeed lead to an isovector central repulsion. However,
the box diagrams (proportional to $g_A^4$) are much more attractive and
overcompensate the expected isovector central repulsion. The isovector central
and isoscalar tensor potentials are shown in Fig.10. We remark that for these
the $1/M$-terms are not negligible. 
\bild{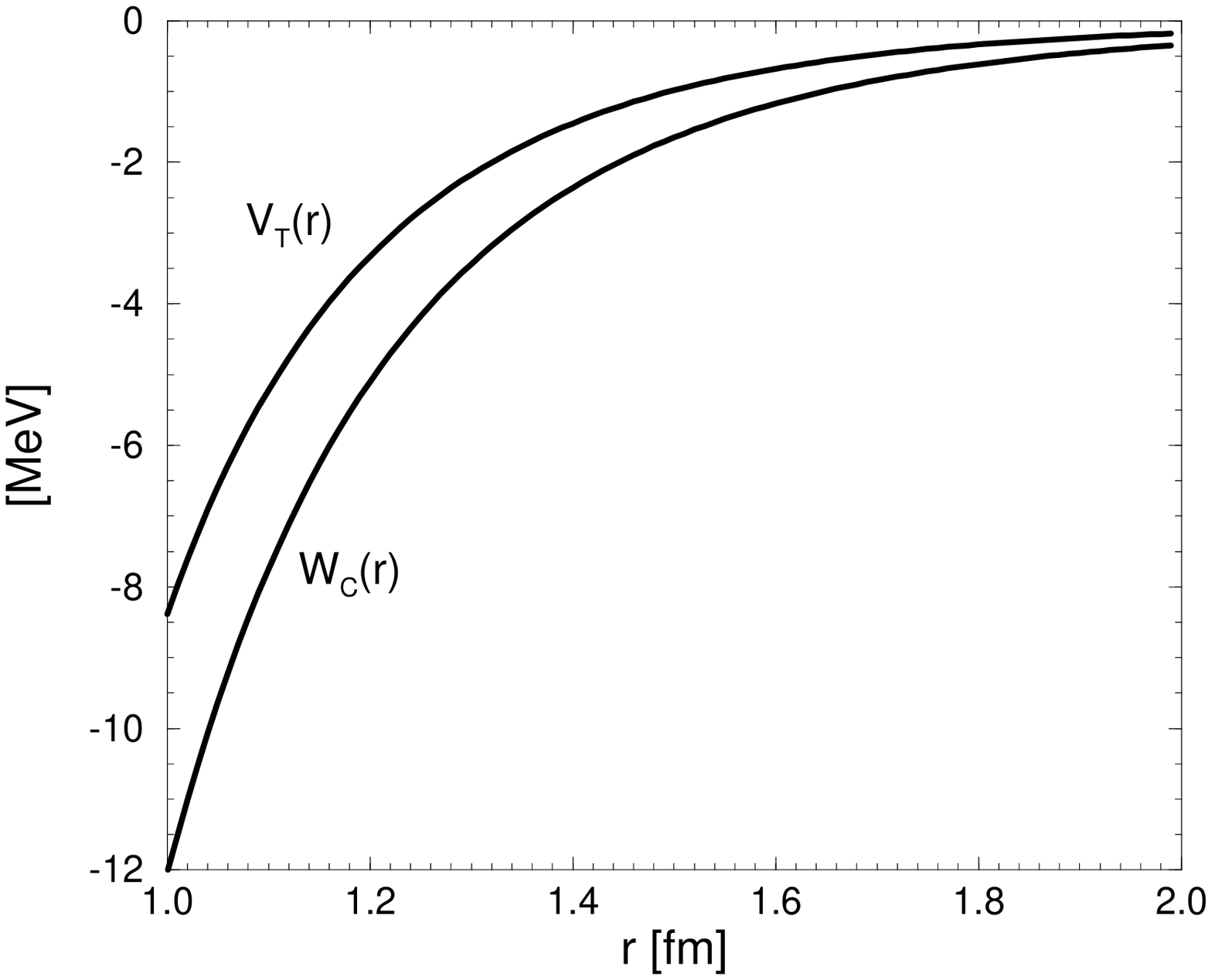}{10}
{\it Fig.10: The isovector central potential $\widetilde W_C(r)$ and the 
isoscalar tensor potential $\widetilde V_T(r)$ generated by irreducible chiral
two-pion exchange in $r$-space.}

\bigskip

At $r=1$ fm they lead to an almost $50\%$ reduction of the leading term, while
at larger distances this reduction gets smaller.

Two-pion exchange spin-spin potentials in coordinate space (accompanying the
operator $\vec\sigma_1 \cdot \vec \sigma_2$) have an analogous spectral 
representation
\begin{equation} \widetilde V_{S}(r)=-{1\over 3\pi^2 r} \int_{2m_\pi}^\infty d
\mu \, \mu^3 \, e^{-\mu r}\, {\rm Im}\, V_T(-i \mu) \,\,. \end{equation}
We find from the chiral two-pion exchange the following isoscalar and
isovector spin-spin potentials,
\begin{eqnarray} \widetilde V_{S}(r) &=& {g_A^4m_\pi\over32\pi^3f_\pi^4\,r^4} 
\bigg\{ 3 x\, K_0(2x) +(3+2x^2)\, K_1(2x ) \nonumber \\ & & -{3\pi e^{-2x}
\over 16 M r} (6x^{-1}+12+11x+6x^2+2x^3) \bigg\} \,\,,
\\ \widetilde W_{S}^{(2\pi)}(r) &=& {g_A^2\over48 \pi^2 f_\pi^4} {e^{-2x} 
\over r^6} \bigg\{ \Big(c_4+{1\over4 M} \Big)  (1+x)(3+3x+2x^2) \nonumber \\ &
& -{g_A^2 \over 16M} (18+36x+31x^2+14x^3+2x^4) \bigg\} \,\,.\end{eqnarray}
Both are positive and thus add to the one-pion exchange spin-spin potential
\begin{equation} \widetilde W_{S}^{(1\pi)}(r) = {g_{\pi N}^2 m_\pi^2 \over 48
\pi  M^2} {e^{- x} \over r} \,\,. \end{equation}
In Fig.11 the spin-spin potentials due to one-pion-exchange (isovector) and
two-pion exchange (isoscalar and isovector) are shown for 1 fm $<r<$ 2 fm. The
isoscalar spin-spin potential is again reduced appreciably by the $1/M$-terms 
in eq.(44). We also note that the terms proportional to $g_A^4$ involving the
modified Bessel-functions $K_{0,1}(2x)$ in eqs.(41,42,44) agree with the
earlier calculation of ref.\cite{brueckner}.  
\bild{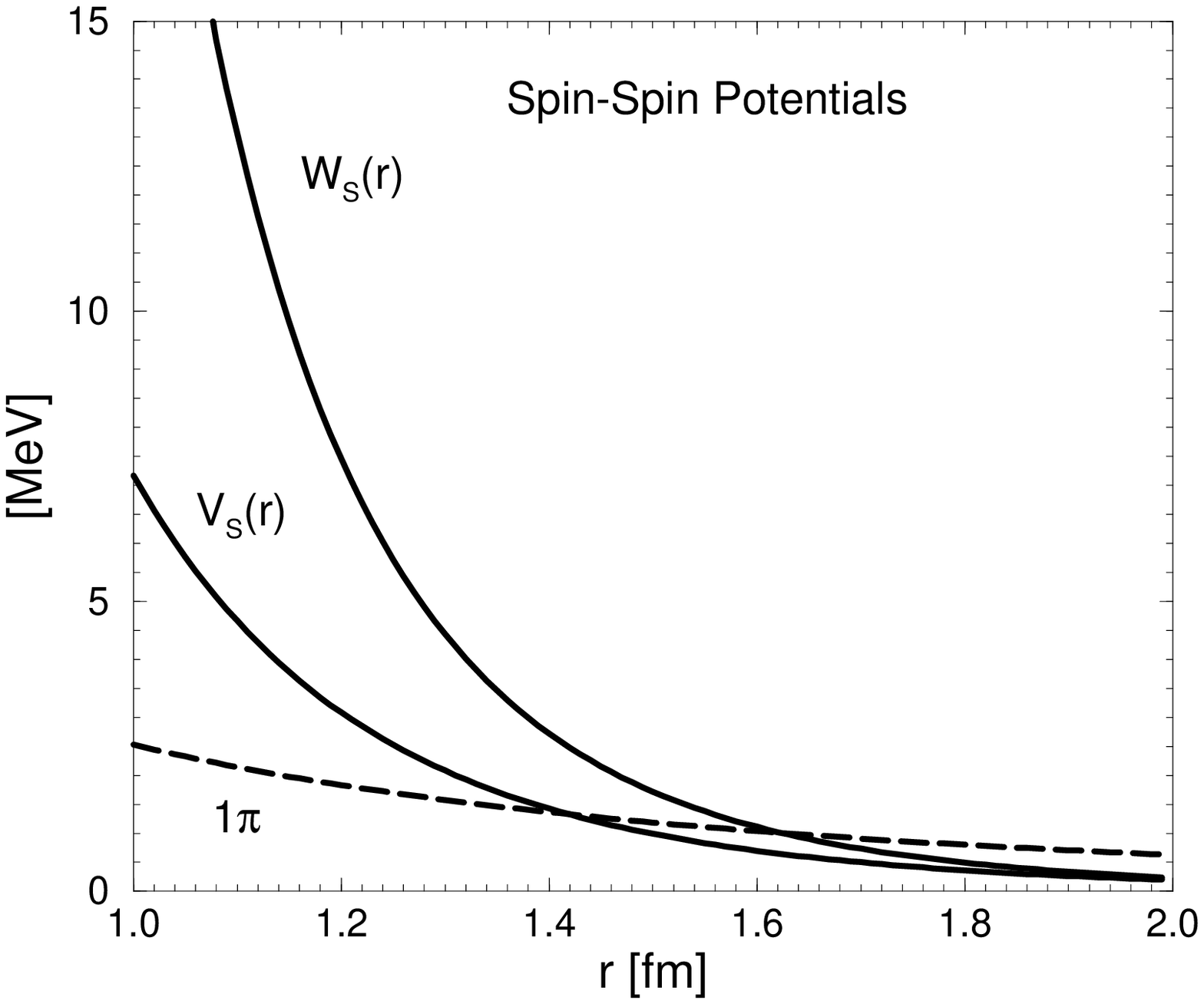}{10}
{\it Fig.11: Spin-spin NN-potentials due to one-pion exchange (dashed line)  
and two-pion exchange (full lines) in $r$-space.}

Spin-orbit potentials in coordinate space (accompanying the standard spin-orbit
operator $-{i\over2} (\vec \sigma_1+\vec \sigma_2) \cdot (\vec r \times \vec
\nabla_r )$) have a spectral representation,
\begin{equation} \widetilde V_{SO}(r) = {1\over \pi^2 r^3} \int_{2m_\pi}^\infty
d \mu \, \mu\,(1+ \mu r)\, e^{-\mu r}\, {\rm Im}\, V_{SO}(-i \mu) \,\,. 
\end{equation}
We find from chiral two-pion exchange the following isoscalar and isovector 
spin-orbit potentials,
\begin{eqnarray}  \widetilde V_{SO}(r) &=& -{3g_A^4\over 64 \pi^2 M f_\pi^4} 
{e^{-2x} \over r^6} (1+x)(2+2x+x^2)\,\,, \\ \widetilde 
W_{SO}(r) &=& {g_A^2(g_A^2-1)\over 32 \pi^2 M f_\pi^4}  {e^{-2x} \over 
r^6}  (1+x)^2  \,\,.\end{eqnarray}
At $r=1/m_\pi = 1.43$ fm these finite range two-pion exchange spin-orbit 
potentials are $\widetilde V_{SO}(1/m_\pi)=- 1.97$ MeV and $\widetilde
W_{SO}(1/m_\pi)= 0.22$ MeV. The isovector spin-orbit potential 
$\widetilde W_{SO}(r)$ is very small and repulsive. However, the isoscalar
spin-orbit potential $\widetilde V_{SO}(r)$  from irreducible two-pion
exchange is quite large and attractive and it turns out to be in very good 
agreement
with the sum of the phenomenological $\sigma(550)$- and $\omega(782)$-exchange
contributions. In Fig.12, these spin-orbit potentials are shown for distances 
1 fm $<r<$  2 fm. Finally, we note that the expressions for the two-pion 
exchange spin-orbit potentials eq.(48,49) agree with the earlier calculation of
ref.\cite{sugawara}. Actually, we agree on all terms with earlier calculations,
except those proportional to $g_A^4/M$ coming from the planar box graph and
contributing to the central, spin-spin and tensor potentials. This type of
relativistic correction is only well defined if the relativistic correction 
to the iterated one-pion exchange is specified (as done here in eq.(24)).  
\bild{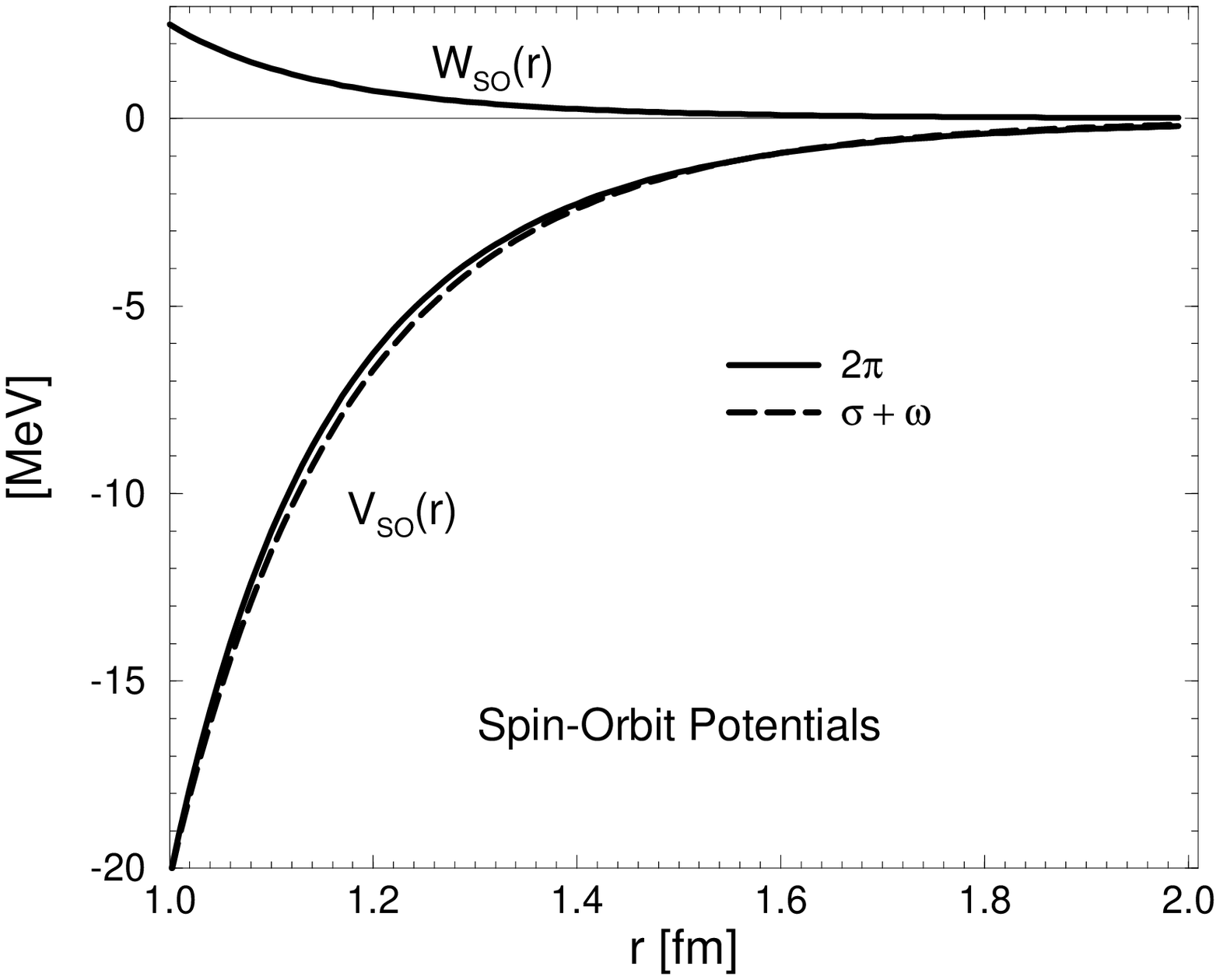}{10}
{\it Fig.12: Isoscalar and isovector spin-orbit NN-potentials of two-pion 
exchange in $r$-space (full curves). The dashed line shows the isoscalar
spin-orbit potential from $\sigma$- and $\omega$-exchange \cite{erwe,bonn}.}

\bigskip
 
As a summary we find,
that all irreducible two-pion exchange NN-potentials are of van-der-Waals
type. They show an asymptotic exponential behaviour, $e^{-2m_\pi r}/r^n\,,n=
{3\over2},2,{5\over2},3,4$, with a decay length $(2m_\pi)^{-1}\simeq 0.7$ fm.
Near the origin $r=0$ they become singular as $r^{-5}$ or $r^{-6}$.   
The strongly attractive parts of the potentials $V_C,\, W_{S}$ and $W_T$ below
$r= 1.5$ fm are responsible for the overestimate of the D-wave phase shifts and
$\epsilon_2$ in Fig.3. Wave-functions of D-states still probe to an appreciable
amount the NN-potentials below $r=1.5$ fm, which in our calculation come out 
too strong in most cases. In order to improve the D-wave observables one has to
add either repulsive zero-range counter terms or an explicit
$\omega(782)$-meson.  

Finally, we have to spell out a warning in order to avoid confusion. The
coordinate space potentials calculated here and the momentum space functions of
section IIB are not related through each other by ordinary Fourier- and inverse
Fourier-transformation. Both of them are too singular at $r=0$ or $ q\to
\infty$ that these transforms would exist. The polynomial pieces $(\alpha+
\beta q^2$) of the momentum space functions generate potentials proportional to
zero-range delta-functions and Laplacians of delta-functions with divergent 
constants. On the other hand the Fourier-transform of a potential as singular
as $r^{-5}$ or $r^{-6}$ near the origin has a divergent polynomial piece in
it. Only in a loose sense, i.e. modulo zero-range $\delta^3(\vec r\,)$-terms
and linear polynomials in $q^2$, are the coordinate space potentials (presented
in this section) and the momentum space functions of irreducible two-pion
exchange  (section IIB) related by Fourier-transformation. For the observables
we have considered in this work, the phase shifts with $L\geq2$ and mixing
angles with $J\geq 2$, these polynomial pieces and delta-function terms do not
play a role at all, as we have mentioned repeatedly.         

\section{COMPARISON WITH ONE-BOSON-EXCHANGE}
A detailed comparison between one-boson exchange and (irreducible)
chiral two-pion exchange in momentum space is not really meaningful because of
the different dependences on the momentum transfer $q$. Nevertheless, we find
it instructive to compare at least the overall strengths of both, i.e. the
values at $q=0$, denoted by an overbar. These values at $q=0$ include also
zero-range effects generated by irreducible chiral two-pion exchange. 

For the isoscalar central part one finds from irreducible chiral two-pion 
exchange and $\sigma+\omega$-exchange at $q=0$, 
\begin{eqnarray}\overline V_C^{(2\pi)}&=& {15g_A^2 m_\pi^3\over 16 \pi f_\pi^4}
\bigg(\, {6\over 5} c_1 -c_3 -{g_A^2\over 64M} \bigg) = 79\,{\rm GeV}^{-2}\,,
\nonumber \\ \overline V_C^{(\sigma)}+\overline V_C^{(\omega)} &=& {g_\sigma^2
\over M_\sigma^2}- {g_\omega^2\over M_\omega^2}= (295-217)\,{\rm
GeV}^{-2}=78\,{\rm GeV}^{-2}\,, \end{eqnarray}
using $g_\sigma^2/4\pi=7.1$ and $g^2_\omega/4\pi=10.6$ \cite{erwe,bonn}. The 
isoscalar central attraction produced by chiral two-pion exchange is stunningly
close to the sum of the individually large phenomenological 
$\sigma$-attraction and $\omega$-repulsion. Note that the isoscalar central
attraction due two-pion exchange is a chiral symmetry breaking effect $\sim
m_\pi^3$. The following interpretation of the numerical result, eq.(50), is 
tempting. In the chiral limit the phenomenological $\sigma$-attraction and
$\omega$-repulsion will cancel each other entirely. On the other hand, it has 
been shown recently \cite{gino} that in this case, $V_C^{(\sigma)}(r)+ V_C^{
(\omega)}(r) \simeq 0$, the so-called pseudospin symmetry observed in nuclear
shell model spectra, can be explained very naturally. This suggests a strong
connection between chiral symmetry of QCD and pseudo-spin symmetry in nuclei. 
Of course, more detailed and quantitative work is necessary to put this
conjecture on a firm basis.
 
We have also compared irreducible chiral two-pion exchange with OBE in other 
channels. No clear relation between OBE ($\sigma, \rho,\omega$) and 
chiral two-pion exchange can be deduced from this comparison. The physics 
behind both is also rather different. Whereas the OBE parametrizes effects of
(strongly) correlated and resonant multi-pion exchange, the chiral two-pion
exchange (to one-loop) involves only soft pions without self-interaction.

\section{SUMMARY}
In this work we have gone one step further in a model independent
description of the low-energy NN-interaction. In addition to the one-pion
exchange, which is known to dominate the high angular momentum partial waves,
we have included two-pion exchange processes based on the most general chiral
effective Lagrangian. We have performed a one-loop calculation of the NN
T-matrix using covariant perturbation theory and dimensional regularization
throughout. We solved the so-called pinch singularity problem occuring in the
planar box diagram by subtracting the iterated one-pion exchange proportional
the nucleon mass $M$. In fact we calculated here all one-loop contributions to
the NN T-matrix at second and third order in the small momentum expansion. 

No pion-nucleon form
factor is generated to modify the point-like one-pion exchange. Restricting 
ourselves to peripheral NN-phases (phase shifts with orbital angular momentum 
$L\geq2$ and mixing angles with $J\geq 2$) the calculation does not involve a
single adjustable parameter. This feature allows for a clean test of chiral
symmetry in the two-nucleon system. Comparing to existing NN phase shift 
analyses we find partial agreement in the D-waves up to $T_{lab}=150$ MeV but
also appreciable deviations.  In particular, we find that the isoscalar central
attraction generated by chiral two-pion exchange is far too strong for
distances 1 fm $<r<$ 2 fm. This points towards the importance of shorter
range effects in the D-waves. 

Increasing the orbital angular momentum $L$ the
agreement between the chiral prediction and the data improves gradually also
for larger energies up  to the $NN\pi^0$-threshold, $T_{lab}=280$ MeV. There is
indeed a kinematical window where chiral symmetry governs the nucleon-nucleon
interaction. Examples for this are the mixing angle $\epsilon_3$ and the
$^1G_4$ and $^3G_5$ phase shifts, where the chiral corrections close the 
sizeable gaps  between the one-pion exchange approximation and the data. In the
high angular momentum  partial waves ($L\geq5$) we find a faster convergence to
the one-pion exchange as obtained in the empirical NN phase shift analyses. 

The calculation presented here incorporates via one- and two-pion exchange the
long and medium long range components of the NN-interaction in a model
independent way. Therefore we propose the chiral NN phase shifts with $L\geq3$
and mixing angles with $J\geq3$ to be used as input in a future NN phase shift
analysis. This way one could find out whether the small differences to existing
phase shift analyses are significant in a direct comparison with NN-scattering
data, namely angular distributions of differential cross sections and
polarization observables. 

We have given explicit expressions for the
isoscalar/isovector central, spin-spin, tensor and spin-orbit potentials in
coordinate space generated by irreducible chiral two-pion exchange. They are of
van-der-Waals type with an exponential asymptotic behaviour, $e^{-2m_\pi
r}/r^n$.  The isoscalar central potential is about twice as attractive as the
one of the $\sigma(550)$-boson in the Bonn potential. Contrary to expectations
the two-pion isovector central potential is attractive, because the
contribution of the box graphs is numerically dominant. 

The one-pion isovector
tensor potential is reduced by the two-pion exchange contribution. The sum of
both is in agreement with the phenomenological Paris potential above $r=1.2$
fm. The two-pion exchange isoscalar spin-orbit potential is surprisingly close
to the sum of the phenomenological $\sigma(550)$- and $\omega(782)$-exchange
contributions.

It is needless to say, that the chiral approach to NN-interaction presented
here can (at the moment) not substitute for the successful phenomenological 
models as a whole. In nuclear physics applications the low partial waves (S, P,
D) are the more relevant ones and chiral symmetry may be of minor importance
for these. A combination of chiral dynamics together with short range
contributions e.g. from $\omega$-exchange might help and will be explored
further.   

\bigskip

\bigskip
\centerline{\bf Acknowledgment}

\medskip

We thank A. Jackson, M. Lutz and R. Machleidt for useful discussions.  

\bigskip

\bigskip

\centerline{ \bf APPENDIX: ANTI-SYMMETRIZED NN T-MATRIX}

\medskip

Even though the (direct) NN T-matrix of the form eq.(4) is sufficient to
calculate all observables it has one deficit. It does not incorporate the Pauli
exclusion principle and therefore it has non-vanishing matrix elements in the
Pauli-forbidden NN-states with even $I+S+L$. This deficit is cured by
subtracting  the T-matrix which results from the exchange of the two out-going
nucleon lines, 
\begin{equation} {\cal T}_{NN} - {\cal A} [{\cal T}_{NN}] \,.\end{equation}
The exchange operation ${\cal A}$ involves a left-multiplication with the
isospin exchange operator $(1+ \vec \tau_1 \cdot \vec \tau_2)/2\,$, a
left-multiplication with the spin exchange operator $(1+ \vec \sigma_1 \cdot
\vec \sigma_2)/2$ and a substitution $\vec p\,' \to -\vec p\,'$ and $z \to -z$
in the scalar functions $V_C(p,z), \dots , W_Q(p,z)$. Evidently, twofold
application of the exchange operation ${\cal A}$ is the identity operation,
${\cal A}\circ {\cal A} = 1\,$, and thus ${\cal T}_{NN}-{\cal A}[{\cal T}_{NN}]
$ is indeed anti-symmetric. The exchange T-matrix ${\cal A} [{\cal T}_{NN}]$
has the same decomposition with respect to the ten isospin-spin operators as
${\cal T}_{NN}$ in eq.(4), only the scalar functions $V_C(p,z),\dots, W_Q(p,z)$
are replaced by different ones, called  ${\cal A}[V_C(p,z)], \dots, {\cal
A}[W_Q(p,z)]$. The latter have the following explicit form (suppressing the
arguments $p$ and $z$),
\begin{eqnarray} {\cal A}[V_C] &=& {1\over4} \Big[V_C +3W_C+3V_{S}+9W_{S}+q^2
(V_T+3W_T)+p^4(1-z^2)(V_Q+3W_Q) \Big]_{z\to -z}\,\,,\nonumber \\ {\cal A}[W_C]
&=&  {1\over4} \Big[V_C -W_C+3V_{S}-3W_{S}+q^2 (V_T-W_T)+p^4(1-z^2)(V_Q-W_Q) 
\Big]_{z\to -z}\,\,, \nonumber \\{\cal A}[V_{S}] &=& {1\over4} \Big[V_C +3W_C-
V_{S}-3W_{S}+q^2(V_T+3W_T)+p^4(z^2-1)(V_Q+3W_Q) \Big]_{z\to -z}\,\,, 
\nonumber \\{\cal A}[W_{S}] &=& {1\over4} \Big[V_C -W_C-V_{S}+W_{S}+q^2
(V_T-W_T)+p^4 (z^2-1)(V_Q-W_Q) \Big]_{z\to -z}\,\,, \nonumber \\ 
{\cal A}[V_T] &=& {1\over 2} \Big[ {z-1\over z+1}(V_T+3W_T) \Big]_{z\to
-z}\,\,, \nonumber \\ {\cal A}[W_T] &=& {1\over 2} \Big[ {z-1\over
z+1}(V_T-W_T)  \Big]_{z\to -z}\,\,,  \nonumber \\ {\cal A}[V_{SO}] &=& {1\over
2} \Big[-V_{SO}-3 W_{SO} \Big]_{z\to -z}\,\,,  \nonumber \\ {\cal A}[W_{SO}] 
&=& {1\over 2} \Big[ W_{SO}-V_{SO} \Big]_{z\to -z}\,\,,  \nonumber \\ {\cal A}
[V_Q] &=& {1\over 2} \Big[  V_Q+3W_Q-{2\over p^2(z+1)}(V_T+3W_T) \Big]_{z\to
-z} \,\,, \nonumber \\ {\cal A} [W_Q] &=& {1\over 2} \Big[V_Q-W_Q- {2\over
p^2(z+1)}(V_T-W_T) \Big]_{z\to -z} \,\,, \end{eqnarray}
with the notation $ [f(p,z)]_{z\to -z}= f(p,-z)$. The matrix elements of ${\cal
A}[{\cal T}_{NN}]$ in the LSJ-basis are readily evaluated replacing $V_C,\dots 
, W_Q$ by  ${\cal A}[V_C],\dots , {\cal A}[W_Q]$ in eqs.(5-9). Making a further
substitution $z\to -z$ and using the property $P_J(-z)= (-1)^J\, P_J(z)$ of the
Legendre polynomials, one finds that the original matrix elements of ${\cal
T}_{NN}$ are reproduced up to an important sign,
\begin{equation} \langle L'SJ |{\cal A}[{\cal T}_{NN}] | LSJ\, \rangle =
(-1)^{I+S+L} \, \langle L'SJ |{\cal T}_{NN} | LSJ\, \rangle\,\,. \end{equation}
The anti-symmetrized T-matrix ${\cal T}_{NN}-{\cal A}[{\cal T}_{NN}]$ has
indeed the desired property, that all its matrix elements in the
Pauli-forbidden NN-states (with even $I+S+L$) vanish identically.  Imposing the
Pauli principle from the very beginning is important if one considers
polynomial counter terms in order to avoid redundance in the low-energy
constants (see e.g. eq.(35) for an S-wave counter term with only two
independent low-energy constants $B,\,B'$).

\end{document}